\documentclass{article}

\usepackage{arxiv}

\usepackage[utf8]{inputenc} 
\usepackage[T1]{fontenc}    
\usepackage{hyperref}       
\usepackage{url}            
\usepackage{booktabs}       
\usepackage{amsfonts}       
\usepackage{nicefrac}       
\usepackage{microtype}      
\usepackage{lipsum}		
\usepackage{graphicx}
\usepackage[square,numbers,comma,sort&compress]{natbib}
\usepackage{doi}
\usepackage{amsmath} 
\usepackage{natbib}

\usepackage{titletoc}

\graphicspath{{figures/}} 

\makeatletter
\newcommand{\mocktitleheader}[1]{%
  \vspace{2em}
  \vbox{%
    \hsize\textwidth
    \linewidth\hsize
    \@toptitlebar
    \centering
    {\LARGE\scshape #1\par}
    \@bottomtitlebar
  }%
  \vspace{1em}
}
\makeatother

\title{Reward function compression facilitates goal-dependent reinforcement learning}

\author{ \href{https://orcid.org/0000-0001-6145-133X}{\includegraphics[scale=0.06]{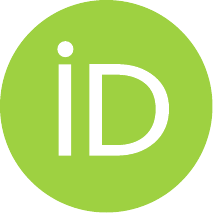}\hspace{1mm}Gaia~Molinaro} \\
	University of California, Berkeley\\
	Berkeley, CA 94704 \\
	\texttt{gaiamolinaro@berkeley.edu} \\
	\And
	\href{https://orcid.org/0000-0003-3751-3662}{\includegraphics[scale=0.06]{orcid.pdf}\hspace{1mm}Anne G.E.~Collins} \\
	University of California, Berkeley\\
	Berkeley, CA 94704 \\
	\texttt{annecollins@berkeley.edu} \\
}

\date{}

\renewcommand{\headeright}{Preprint, 2025}
\renewcommand{\undertitle}{}
\renewcommand{\shorttitle}{Reward function compression facilitates goal-dependent reinforcement learning}

\hypersetup{
pdftitle={Reward function compression facilitates goal-dependent reinforcement learning},
pdfsubject={q-bio.NC, q-bio.QM},
pdfauthor={Gaia~Molinaro, Anne G.E.~Collins},
pdfkeywords={First keyword, Second keyword, More},
}

\begin{document}
\maketitle

\begin{abstract}
Humans can uniquely assign value to novel, abstract outcomes to support reinforcement learning. However, this flexibility is cognitively costly and reduces learning efficiency. We propose that goal-dependent learning initially relies on capacity-limited working memory. With consistent experience, learners create a ``compressed'' reward function -- a simplified goal rule -- that transfers to long-term memory for a more automatic evaluation upon receiving feedback. This automaticity frees working memory resources, thereby boosting learning efficiency. Across six experiments, we demonstrate that learning is impaired by the size of the goal space but improves when this space allows for compression. Additionally, faster reward processing correlates with better learning. Although the algorithmic details remain to be established, our behavioral results and computational models suggest that efficient goal-directed learning relies on compressing complex goal information into a stable reward function. These findings illuminate the cognitive mechanisms of intrinsic motivation and can inform behavioral interventions supporting human goal achievement.
\end{abstract}


\section*{Significance statement}
From completing crosswords to achieving a career ambition, people pursue a wide variety of goals. However, this flexibility is often cognitively effortful, which can make it difficult for individuals to learn how to achieve their objectives. We propose that the brain manages this trade-off by initially relying on a fast, but expensive working memory system; with experience, it attempts to ``compress'' complex goal information into a simplified rule that is applied automatically, thereby freeing up mental resources and boosting learning. Our experiments validate this model, revealing a key process in human motivation that can inform strategies for more effective learning and goal attainment.

\section*{Introduction}

Reinforcement learning is a powerful process by which agents learn to make decisions based on experienced rewards and punishments, which has seen broad applications across neuroscience, psychology, and machine learning \cite{EcksteinAndCollins2021}. The study of animal reinforcement learning (RL) has predominantly focused on two outcome categories: primary rewards, which satisfy basic physiological needs (e.g., food, water), and secondary rewards, which derive their value through learned associations with primary rewards (tokens, money, numeric points \cite{Skinner1984}). This focus has advanced our understanding of RL behavior and its neural basis in dopaminergic signaling \cite{Niv2009}. However, primary and secondary rewards do not adequately capture the complexity of human reward processing \cite{DanielAndPollmann2014}, especially in contexts involving abstract goals and deliberate planning \cite{DiukEtAl2013}. For example, people often find activities rewarding in the absence, if not at the expense, of external incentives \cite{BlainAndSharot2021}. Even when the benefits of engaging in an activity are clear, people can attribute value to its successful completion a priori, without needing to directly associate new outcomes with established rewards. Indeed, humans can flexibly imbue novel outcomes with value and use them to guide learning through a process that recruits subcortical reward circuits \cite{McDougleEtAl2022, MolinaroAndCollins2023Hacks}. Together, these findings suggest that humans use a third class of rewards: flexible, goal-dependent outcomes that derive value from current objectives rather than learned associations \cite{MolinaroAndCollins2023GoalCentric}. 

Goal-dependent reinforcement learning implies the construction of a goal-dependent reward function, i.e., a mapping between an outcome space and scalar rewards. While individuals show remarkable flexibility in creating ad hoc reward functions according to current goals, the underlying process is often inefficient \cite{McDougleEtAl2022}. This is illustrated, for example, by the discrepancy between the high frequency at which people set New Year's resolutions and the low rate at which they complete them \cite{WoolleyEtAl2025}. However, the precise cognitive mechanisms by which humans align their value system to current goals -- and which aspects make such alignment hard to maintain -- remain uncharted. Identifying the key players involved in goal-dependent learning and their respective weaknesses could inform behavioral strategies that help people achieve their objectives in both personal and educational settings. 

Setting up custom reward functions likely involves the interplay of multiple cognitive systems \cite{McDougleEtAl2022}. The existence of interactions between RL and executive functions is now well-established \cite{RmusEtAl2021, YooAndCollins2022}. In particular, WM can act as a source of information for RL, affording faster or more efficient learning \cite {CollinsAndFrank2012, CollinsAndFrank2018}. WM has also been proposed to support the verbalization of critical task components \cite{YooEtAl2023, RmusEtAl2023} and the representation of states, actions, and rewards over which RL operates \cite{HamEtAl2025}. However, much less is known about how exchanges between RL and WM support goal-dependent reward functions, and how they might be limited.
Building on existing work \cite{McDougleEtAl2022}, we designed a task where participants had to learn which responses led to desired outcomes, but using abstract images labeled as ``goal''/``nongoal'' as feedback instead of standard numeric points (Figure \ref{fig:task_structure}). We replicated our finding that people can indeed imbue abstract, novel stimuli with reinforcing properties, but that learning from goal-dependent outcomes resulted in worse learning performance (\cite{McDougleEtAl2022, MolinaroAndCollins2023Hacks}; Supplementary information -- Experiment 3), highlighting the trade-offs between flexible goal-dependent value attribution and learning efficiency. This impairment persisted (albeit substantially smaller) when goal images were kept consistent for the entire duration of the task (\cite{MolinaroAndCollins2023Hacks}, Supplementary information -- Experiment 4). However, evidence for the precise reason behind such performance improvements remained inconclusive. 

\begin{figure}
\centering
\includegraphics[width=0.5\linewidth]{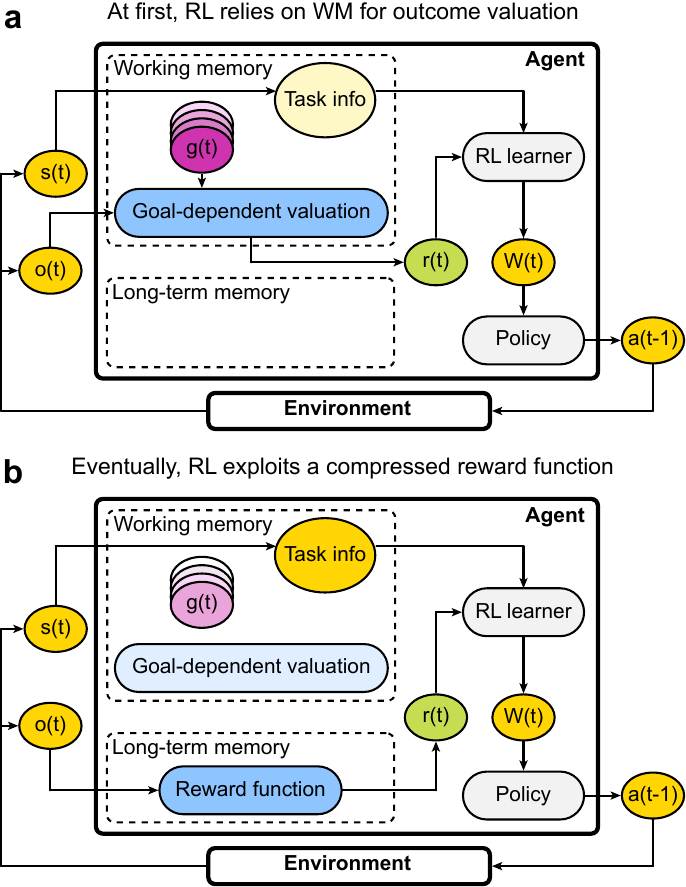}
\caption{\textbf{Conceptual model of goal-directed reinforcement learning with reward function compression.} In our proposed conceptual model, we break down goal-dependent learning into several parts. Combined with other learning components, an RL learning algorithm informs value estimates $W$ on a given trial $t$, which are entered into a policy. The agent selects an action $a$ through a joint policy, to which the environment responds by outputting outcome information $o$ and a new state $s$. A working memory (WM) component organizes and maintains task-relevant information (here, broadly intended and dubbed the ``task information''), used to support the reinforcement learning (RL) process. \textbf{a} Initially, value attribution to $o$ is operated by WM, which maintains the current goal and supports its matching to the experienced outcome to generate a reward $r$ used by the RL agent. WM also holds a trace of recently experienced goals. 
\textbf{b} If goals share enough similar features, a compressed reward function can be created, which replaces active goal maintenance and can, over time, be transferred to long-term memory storage. Eventually, the compressed reward function can be directly employed to assign value to experienced outcomes, sparing resources for other WM tasks.   
}
\label{fig:schematic}
\end{figure}

Here, we propose a dynamic model of goal-dependent reward processing to capture the observed trade-offs between flexibility and learning efficiency. 
As a real-life example, consider learning chess: a novice might initially hold specific winning positions in mind (``queen on D5''), then compress these into simple tactical rules (``control the center squares''), and eventually apply such patterns automatically without conscious effort, freeing mental resources for strategic planning.
In our conceptual model, a working memory (WM) component enables us to instantly attribute goal-contingent rewarding properties even to abstract, novel goals. This allows for almost complete freedom in defining what counts as a ``reward'' (including usually unrewarding outcomes \cite{PanEtAl2025}), which can then be used as a signal for learning (Figure \ref{fig:schematic}a). However, such freedom comes at a cost, as it requires active maintenance of a goal \cite{Baddeley1998}, using up resources otherwise utilized for the learning process itself \cite{Oberauer2019}. 
Information stored repeatedly in WM transfers to long-term storage that resists cognitive load \cite{CarlisleEtAl2011, Logan1988, ShiffrinAndSchneider1977, WoodmanEtAl2006}. Consequently, we propose that pursuing similar goals consistently alleviates the burden on WM through the creation of a ``compressed'' reward function from high-dimensional stimuli, which contains the minimum amount of information required to recognize successful outcomes as such while trimming redundant details away (e.g., ``all goal images are purple''). Such a lightweight reward function can eventually be transferred to longer-term memory storage, from where it may easily be retrieved without impinging on other executive functions required to support learning through RL or WM (\cite{RmusEtAl2021, YooAndCollins2022, YooEtAl2023} Figure \ref{fig:schematic}b). 

While we remain agnostic with regard to its algorithmic implementation in the brain and formal instantiations of reward function compression itself, our model makes specific predictions, which we test in this study.
If WM supports goal evaluation, then learning should worsen as we increase the number of goals to build a reward function for across the learning problem (i.e., WM load).
In Experiment 1, we manipulate the number of unique goals participants need to learn from, and observe that learning parametrically decays.   
The model also predicts that performance will benefit from repeated encounters with the same set of goals when it either naturally fits within WM capacity or can be compressed to do so. Only then can the transfer to long-term memory occur, freeing up executive resources to support learning. In Experiment 2, we keep the number of goals constant across blocks, but vary whether the goal space can be compressed along a subset of its dimensions. We find that learning performance significantly improves when people are able to build compressed reward functions, compared to when they are not. 
In Supplementary Experiments 3-6, we eliminate alternative mechanisms by which reward functions may be set following instruction, such as cost-free top-down value attribution or cached estimates of reward that are acquired iteratively. 
While no single experiment eliminates all competing hypotheses at once, the study as a whole tells a coherent story. Taken together and paired with computational modeling, our behavioral findings support reward function compression as a possible mechanism for efficient goal-dependent reinforcement learning. 

\section*{Results}

\subsection*{Experimental design}
\begin{figure}
\centering
\includegraphics[width=0.65\linewidth]{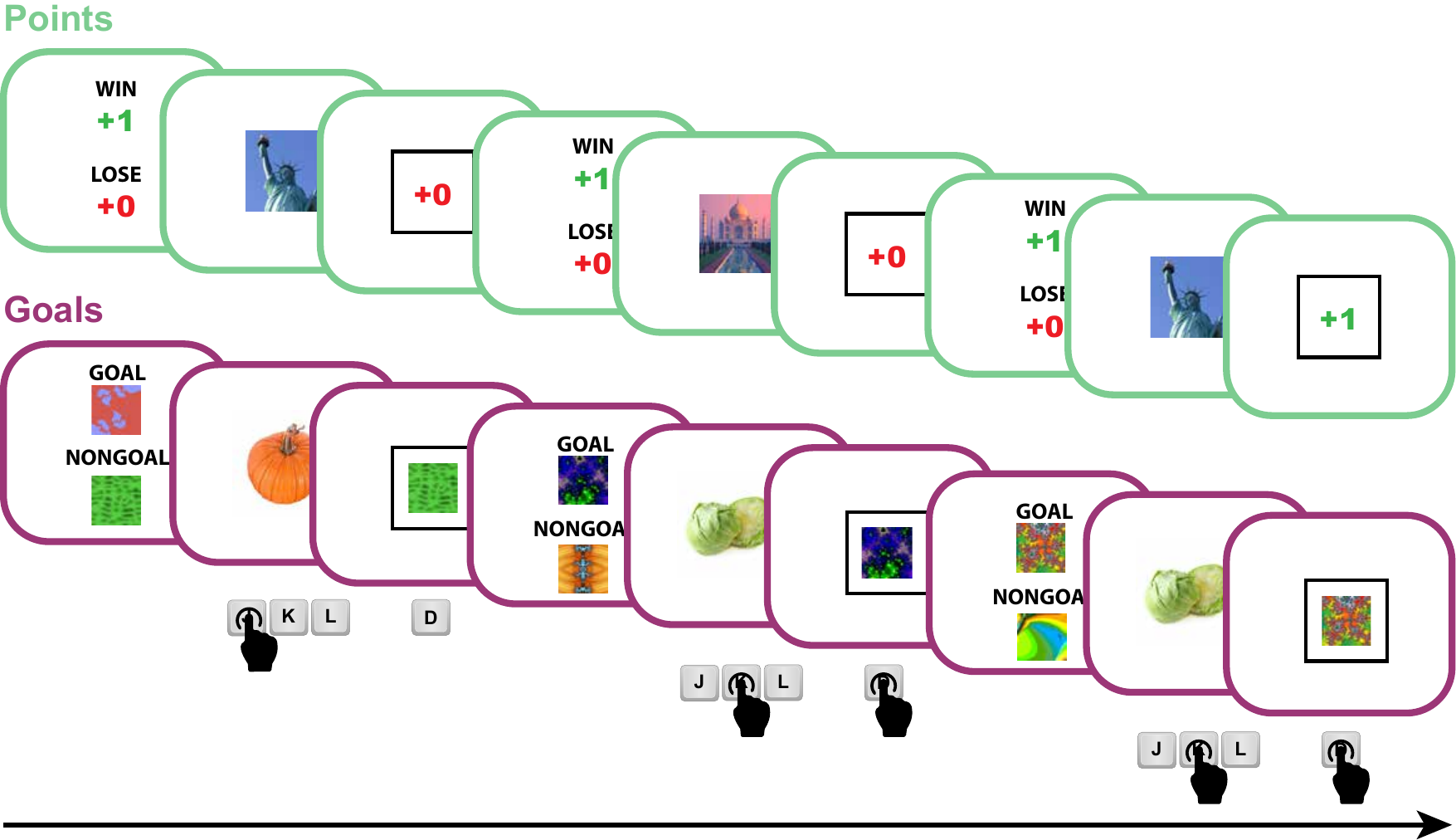}
\caption{\textbf{Three example trials for each within-subjects condition.} Participants learned the mapping between each of six images and a stimulus-specific, deterministic correct response across several iterations. Within Points blocks, feedback was provided in the form of ``+1'' or ``+0'' messages. Within Goals blocks, ``goal'' and ``nongoal'' images (independent of stimuli) were used as feedback.  Participants were also instructed to ``collect'' desirable outcomes by pressing a separate key upon obtaining them while ignoring undesirable outcomes.}
\label{fig:task_structure}
\end{figure}

All experiments share the same fundamental structure, which we adapted from existing tasks \cite{CollinsAndFrank2012, McDougleEtAl2022} where participants had to learn, by trial and error, the correct association between stimuli and specific keys on their computer keyboard (Figure \ref{fig:task_structure}). Both Experiment 1 and Experiment 2 consisted of six blocks, divided into interleaved ``Points'' and ``Goals'' blocks (two and four of each, respectively) in a pseudo-randomized order. Six stimuli were presented in each block and repeated 12 times in an interleaved fashion, for a total of 72 trials. Compared to the original RLWM task \cite{CollinsAndFrank2012}, only having blocks with six stimuli in each allowed us to emphasize the RL component relative to value updates directly informed by WM. In the Points blocks, which reflected the structure of a standard RL task and served as a baseline, participants received numeric points as feedback for their responses to the stimuli. Each trial started with a fixation screen (0.5 s), followed by the presentation of ``+1'' and ``+0'' messages indicating desirable and undesirable outcomes, respectively (1.8 s). Participants then saw a stimulus image and were prompted to press one of three keys on their keyboard (1.5 s). Upon responding, they received deterministic feedback\footnote{Although feedback was deterministic, participants were in a learning regime for the majority of the task, such that reward prediction errors did not saturate until the very end, if ever (Figure \ref{fig:behav_model_results}c,g).} showing ``+1'' or ``+0'' messages based on the accuracy of their response (1.5 s) or a warning message if they failed to respond on time. They were instructed to ``collect'' desirable outcomes by pressing a separate key within the allotted time while ignoring undesirable outcomes (as in a Go/No-go task). Below, we refer to the latter stage of each trial as ``reward collection''. The Goals blocks followed a similar structure but used abstract fractal-like images instead of numeric points. In each trial, before the stimulus was shown, a pair of fractal images was displayed with the labels ``Goal'' and ``Nongoal'' to indicate the desired outcome. Participants needed to encode and recall which images represented goals to learn the correct stimulus-action mappings, then collect ``goal'' outcomes on each trial. Correct action selection for each stimulus image deterministically led to the presentation of the current trial's goal image. 

\subsection*{Experiment 1: Goal-dependent value attribution is impacted by load}

Our conceptual model (Figure \ref{fig:schematic}) proposes that participants will, over time, try to infer a simplified reward function that can be applied to the presented outcome, as opposed to remembering each goal image separately and later comparing it to the experienced outcome. 
The easier the reward function compression process, the faster goal-dependent reward functions can be ascribed to long-term memory, freeing up cognitive resources for the learning process itself. 
If only one goal is being pursued across a set of trials, it is easy to compress reward information from the complex fractal goal image to one or a few defining attributes (e.g., color, pattern, etc.) that distinguish it from the nongoal image. When multiple goals are present across trials, such an operation increases the load on WM and thus becomes harder. 
Therefore, we predicted that modulating the number of goal images in each block would result in performance gains or losses.
Returning to our chess example, this would be akin to working toward a single target position across many games vs. juggling a repertoire of several distinct target positions. The latter is harder because resources needed for learning must be spared for active goal maintenance. 

In each of the Goals blocks of Experiment 1, outcome images belonged to a set of one, two, three, or four pairs, such that each fractal image pair appeared 12, six, four, or three times per stimulus (Figure \ref{fig:behav_model_results}a). Block order was pseudo-randomized. We additionally ensured that a Points block appeared within the first three and last three blocks. We expected learning performance to be best in the block with a single goal, and worst in the block with four different goals to pursue, with intermediate learning efficiency in the conditions in between. 
Importantly, conditions did not differ in the amount of information participants had to memorize on a trial-by-trial basis, since only a single goal/non-goal pair was relevant in each trial. Thus, the difference in load was only effective across trials, in keeping with the assumption that participants would consider information from the entire goal space and try to build an overall goal-dependent reward function.
Sixty-seven participants completed Experiment 1, 12 of whom did not match our inclusion criteria (see \hyperref[sec:methods]{Methods}). We analyzed data from the remaining 54 participants (74\% female, ages 18-25, M = 20.13 $\pm$ 1.48).

Learning performance was higher for Points, compared to Goals trials -- even when compared to the easiest, single-goal block (t(53) = 3.74, p = 0.001), highlighting the cost of learning from goal-dependent outcomes even in a simple setting and confirming our previous findings \cite{MolinaroAndCollins2023Hacks}.
To compare performance across blocks with one, two, three, or four pairs of goal/nongoal outcomes, we ran a logistic mixed-effects model predicting average learning performance from the number of outcome image pairs in each Goals block, with random intercepts for participant number. 
Performance decreased with the number of different outcomes in a block ($\beta$ = 0.073 $\pm$ 0.02, 95\% CI = [0.04, 0.01], p $<$ 0.001; Figure \ref{fig:behav_model_results}b-c). 
By contrast, reward collection accuracy was similar in Goals (M = 3.86 $\pm$ 0.08) and Points trials (M = 3.78 $\pm$ 0.09; t(53) = -0.81, p = 0.424, BF10 = 0.20 i.e., moderate evidence for the null hypothesis) and did not differ across Goals blocks ($\beta$ = -0.01 $\pm$ 0.06, 95\% CI = [-0.12, 0.10], p = 0.820; Supplementary Fig. 1), showing that the load manipulation did not impact outcome identification and goal maintenance within trials.  

In sum, modulating the number of goal/nongoal image pairs from one to four resulted in parametric impairments in learning. This sort of phenomenon, with relatively constrained ceiling-floor effects and proportionally increasing performance in between, is a hallmark of WM \cite{LuckAndVogel1997}, suggesting a similar executive function might be at play in the reward function compression process. 
However, it is also possible that the observed learning impairments were caused by interference, occurring more frequently in blocks with more unique goal images. 
To disentangle the two hypotheses, we devised Experiment 2. 

\subsection*{Experiment 2: Reward function compression facilitates goal-dependent learning}
\begin{figure*}
\centering
\includegraphics[width=1\textwidth]{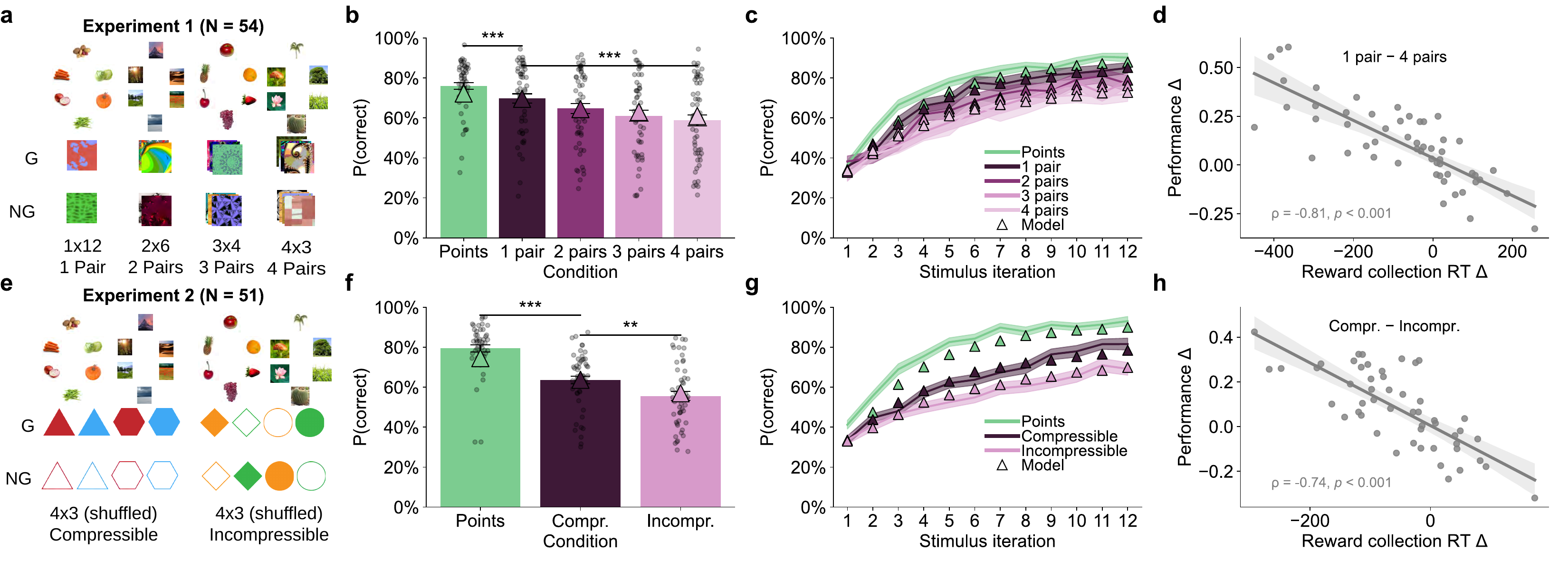}
\caption{\textbf{Experiment manipulations, learning performance, and reward collection reaction times (RTs; in ms).} \textbf{a, e} Goal block manipulations. G = Goal image, NG = Nongoal image. \textbf{b, f} Average choice accuracy across conditions. \textbf{c, g} Learning performance over stimulus iterations. \textbf{a, h} Correlation between differences in learning performance and reward collection RTs. The top row shows results for Experiment 1, the bottom row for Experiment 2. Bars, dots, and solid lines indicate participants' behavior. Triangles indicate simulations from the best-fitting model, where learning rates were informed by reward collection RTs. Error bars and shading show the S.E.M. *** p $<$ 0.001, ** p $<$ 0.01.}
\label{fig:behav_model_results}
\end{figure*}

Our model predicts that the mere repetition of goals is not always sufficient for performance improvements: the goals must be structured in a way that allows for meaningful compression. If participants were simply benefiting from seeing the same goal images repeatedly (as might happen in basic interference accounts), then any repeated goal set should show improvement. However, if our reward function compression hypothesis is correct, performance should improve only when the goal space can be reduced to a simple, memorable rule.
Experiment 2 tests this prediction by holding both the number of unique goal images and their frequency of presentation constant while manipulating whether the underlying structure of the goal space permits reward information compression through dimensionality reduction.
To achieve this, we replaced fractal images with simpler geometric figures that varied along three different binary axes within each block (Figure \ref{fig:behav_model_results}e). This resulted in eight figures per block, each having one of two shapes (e.g., square or triangle), colors (e.g., purple or yellow), and fills (full or empty). In ``Compressible blocks'', a single feature was sufficient to separate goal from nongoal images, while other features were irrelevant (e.g., all goal images were full; classification type I in \cite{ShepardEtAl1961}). In ``Incompressible'' blocks, all three features had to be taken into account to determine whether an outcome was a goal or a nongoal image (e.g., goal images were a full orange cross, a full green circle, an empty green circle, or an empty orange circle; classification type VI in \cite{ShepardEtAl1961}). 
The same rules governed blocks of the same type for a given participant, but rule assignment was pseudorandomized across individuals.
Going back to the chess analogy, the two Goals block types are like learning the same number of target positions, but differ in whether all winning positions can be generalized via common features (e.g., controlling the center) or depend on the conjunction of several features, leaving the player to memorize each target board in full.
Seventy-one participants completed Experiment 2. 
After excluding 20 who failed to meet our inclusion criteria, we analyzed data from the remaining 51 participants (80\% female, ages 18-32, M = 20.48 $\pm$ 2.27).

As predicted, learning accuracy was significantly higher in Compressible than Incompressible blocks (t(50) = 3.31, p = 0.002; Figure \ref{fig:behav_model_results}f-g). Since the number of occurrences of each goal outcome was identical across the two block types, these results cannot be attributed to differences in goal outcome repetitions (as could have been the case for Experiment 1). 
Goals block type differences in reward collection accuracies were significant (t(50) = 8.62, p $<$ 0.001; Supplementary Fig. 1). However, they were not significantly correlated with corresponding differences in learning performance ($\rho$ = 0.17, p = 0.226). These results exclude failures in goal identification as a possible explanation for the effect. 
Moreover, despite new learning, reward collection RTs decreased from the first set of Goals blocks (M = 422.18 ± 13.92) to the second (where goal/nongoal images were reused; M = 316.06 ± 15.9, t(50) = 8.0, p < 0.001). This RT decrease was sharper for the Compressible, compared to the Incompressible condition: a linear mixed-effects model predicting reward collection RTs showed a significant interaction between Goals condition type and experiment phase (early vs. last; $\beta$ = -13.58 $\pm$ 6.14, 95\% CI = [-25.62, -1.53], p $<$ 0.001), controlling for stimulus iteration and its interaction with condition type. Similar findings applied to learning performance increases across phases (early learning accuracy M = 0.53 $\pm$ 0.02, late learning accuracy M = 0.66 $\pm$ 0.02; t(50) = -9.45, p $<$ 0.001; Goals condition type by experiment phase interaction $\beta$ = -0.03 $\pm$ 0.01, 95\% CI = [0.01, 0.04], p $<$ 0.001). Together, these results suggest that learning benefits could not be fully attributed to more habitual choice behavior (Supplementary Fig. 2).
 
Together, these results suggest that, although they were never instructed to try to learn mappings between goal outcome features and reward, participants were able to leverage structure in the condition with ``compressible'' goal outcomes, which ultimately benefited their learning. 

\subsection*{Reward processing correlates with performance}

During the task, participants were instructed to ``collect'' positive outcomes (either points or goal images) by pressing a specific key while ignoring negative outcomes, a phase we refer to as ``reward collection'' (Figure \ref{fig:task_structure}). 
In this section, we explore reward collection reaction times (RTs; Supplementary Fig. 1) as a behavioral marker of reward function compression.
As participants switch from controlled to more automatic reward processing, their reward collection should become faster. The quicker an individual can recognize and react to an outcome as desirable, the less cognitive effort is presumably expended on this evaluative step. This, in turn, could free up cognitive capacity for the main task of learning new stimulus-response associations. Therefore, we expected individual differences in reward collection RTs across conditions to correlate with corresponding differences in learning performance. By contrast, faster reward collection RTs due to motor habituation, general arousal, or automatic valuation would not systematically differ across the various conditions.

In each experiment, average block-wise individual differences between Points and Goals learning accuracy were strongly correlated with corresponding differences in median log reward collection RTs (Supplementary Fig. 3). Similarly, performance differences between pairs of Goals conditions within the same experiment correlated with corresponding reward collection RT differences (Figure \ref{fig:behav_model_results}d, h, Supplementary Fig. 3). We sought to ensure that this finding was not driven by differences in engagement or difficulty across conditions. To this end, we reasoned that, if a third, non-specific factor jointly drove block-level changes in choice accuracy and reward collection RTs, we should also expect it to drive choice RTs, and thus a similar correlation between performance and RT. Contrary to this prediction, we found weak or no correlation between choice RT and performance differences across conditions of Experiments 1-2 (Supplementary Figs. 6-7). This shows that block-level reward collection RT is uniquely predictive of accuracy. This finding is also counter to the competing hypothesis that goal maintenance impacts choice through a dual-tasking effect, whereby the cognitive load of maintaining the goal to interpret goal-dependent feedback would interfere with and slow down the choice process.

This reliable -- albeit correlational -- link between reward collection RTs and learning success is difficult to reconcile with simpler alternative models. For instance, an immediate ``top-down'' value attribution process would likely be uniformly efficient, leading to consistent reward collection speeds. Similarly, ``bottom-up'' explanations that lack an abstraction mechanism would struggle to account for these results -- especially in Experiment 2, where goal outcomes were equally frequent but more or less diverse. 
By contrast, the observed pattern of results strengthens the case for an active reward compression process supporting goal-dependent learning, whereby WM information can be transferred to longer-term storage and be applied without interfering with other executive functions. To test the internal validity of our model, we turn to computational models, leveraging reward collection RTs as a signature of reward function compression. 

\subsection*{Computational models leveraging signatures of reward function compression accurately recover human behavior}

To gain a closer understanding of the latent processes guiding learning and decision-making in our task, we fit a set of candidate models that shared a basic RL architecture \cite{SuttonAndBarto2018}. 
Details about computational modeling methods are provided in the \hyperref[sec:methods]{Methods} section.
In brief, the value of each stimulus-action pair was updated on each trial through the delta rule \citep{RescorlaAndWagner1972}. Actions were selected via a noisy softmax policy \cite{Luce1959}. This basic RL component had a learning rate $\alpha$, a learning rate multiplier for negative outcomes $\zeta$, a fixed inverse temperature $\beta$ = 25, and a free noise parameter $\epsilon$. We note that this RL component is meant to capture the reward-dependent learning process, with no claim that it maps onto specific underlying neural mechanisms \cite{EcksteinAndCollins2021, Collins2026}; indeed, we expect that this RL component additionally reflects contributions of working memory to learning \cite{Collins2018}.
All models also included additional mechanisms, such as forgetting and stickiness, which we found to improve fit. In particular, forgetting proxies RL-WM mixtures that could not be disentangled due to the fixed set-size design of the experiment. 
An additional learning process tracked stimulus-action associations in an outcome-insensitive, habit-like manner \cite{Collins2026}. This component had a separate learning rate $\tau$ for outcome-insensitive value estimates, and a parameter $\kappa$ that defined the mixture of outcome-sensitive and outcome-insensitive values the RL learner would take into account. 

This model, in its base form, best captured the overall learning dynamics, but could not support differences between Points and Goals blocks or within-Goal block sub-conditions. To remedy this, we expanded the base model with separate free parameters for each condition of interest, one mechanism at a time. This resulted in a new family of models where Points trials and each condition of Goals blocks had a specific learning rate, noise parameter, or stickiness bias, etc. Among these, the winning model was the one with separate learning rates $\alpha$ for each type of block (Supplementary Fig. 8). In line with previous studies \cite{McDougleEtAl2022}, this result suggests that the value updating process (and not, for instance, the choice selection aspect) was the most negatively impacted by goal-dependent reward attribution. 

Treating reward collection RTs as proxies for reward processing efficiency (potentially gained through long-term memory transfer), we also built a family of models where one of the parameters was directly influenced by the block-wise median (log) reward collection RT for the corresponding condition (i.e., Points, or a specific Goals sub-condition; see \hyperref[sec:methods]{Methods} for details) to a degree defined by the parameter $\omega$. 
This model design rests on the assumption that faster RTs indicate more efficient processing and that, if reward-processing gains free up shared resources, they should relate to specific benefits in learning. In contrast, if learning improvements were unrelated to reward processing, linking the two systems would not capture behavior more accurately than a model with parameters independent of reward collection RTs.
By applying this method to the learning rate, we were able to capture participants' behavior (Experiment 1: Figure \ref{fig:behav_model_results}b-c; Experiment 1: Figure \ref{fig:behav_model_results}f-g) at least as well, and more parsimoniously than the model with free parameters for each experimental condition (Experiment 1: $\Delta$AIC = 0.91, Wilcoxon's Z(53) = 564.0, p = 0.124; Experiment 2: $\Delta$AIC = 5.95, Z(50) = 225.0, p $<$ 0.001; Supplementary Fig. 9. This finding provides formal support for our hypothesis that, when outcome evaluation (proxied by reward collection RT) is efficient, more cognitive resources can be allocated to update stimulus-action associations from feedback.  
Moreover, the model with learning rates conditioned on reward collection RTs provided a better fit and posterior predictive checks to participants' data than a model where the choice noise parameter $\epsilon$ was altered by reward collection RTs (Experiment 1: $\Delta$AIC = 8.24, Z(53) = 84.0, p $<$ 0.001; Experiment 2: $\Delta$AIC = -8.53, Z(50) = 136.0, p $<$ 0.001; Supplementary Figs. 8, 13), showing that the bottleneck in goal-dependent learning occurs at the value-updating stage.
We note that these results cannot simply be attributed to a shared variance in the learning process captured by reward collection RT. To illustrate this point, we also fit a model where the learning rate is affected by average \textit{choice} RT and show that it does not explain behavior as accurately (Experiment 1: $\Delta$AIC = -5.32, Z(53) = 302, p $<$ 0.001; Experiment 2: $\Delta$AIC = -7.10, Z(50) = 238, $<$ 0.001; Supplementary Fig. 8, S14). 

\subsection*{Supplementary experiments}
We conducted four supplementary control experiments (Supplementary information -- Supplementary experiments, Supplementary Fig. 3) that provide further support to our model by establishing basic phenomena or eliminating some key alternative explanations. 
Experiments 3 and 4 are re-analyses of previous studies \cite{MolinaroAndCollins2023Hacks}. 
In Experiment 3, goal images changed on each trial, which made reward function compression impossible and forced participants to attribute value in a one-shot manner. This experiment provides a close replication of \cite{McDougleEtAl2022} and shows that goal-dependent learning is more effortful than learning from standard reinforcers. Here, we also find that simple lapses in goal outcome encoding or recognition do not explain impairments in learning performance. This is consistent with the results presented above and suggests that while goal maintenance in itself is efficient, it is a costly process that can utilize resources otherwise allocated to learning.
In Experiment 4, goal images were kept consistent across blocks, facilitating reward function compression. We find that, while goal-dependent learning is initially impaired, it eventually reaches similar efficiency levels to learning from standard rewards. We attribute such dynamic changes to a transfer of information from WM to longer-term memory storage. Confirming this hypothesis, we find that individual preferences for goal, over nongoal outcome images -- initially neutral -- become stronger over time. 
Although Experiment 2 already showed that repeating goals is not sufficient to achieve performance savings, we provide a more controlled test for this hypothesis in Experiment 5, where we directly manipulate the number of goal outcome repetitions within a block. We find no performance gains for more frequent goal outcomes, which we attribute to the fact that, when the goal space is beyond WM capacity, there is no opportunity for reward function compression, meaning WM is continually burdened. 
In Experiment 6, we directly address the hypothesis that goal-dependent value attribution is purely done in a one-shot manner, as opposed to being sensitive to the previous history of goal setting. To this end, we manipulate whether goal/nongoal labels remain consistent across trials, controlling for the number of unique goals and how often each is presented, and find that learning is specifically impaired by inconsistent goal/nongoal image labeling across trials. This result suggests that goal-dependent learning takes advantage of repeated goal instances, which we attribute to reward function compression and transfer to longer-term memory storage, and which can make resources available for other online computations. 

\section*{Discussion}

From mundane activities to long-term ambitions, humans are remarkably capable of setting and pursuing a vast range of goals \cite{ColasEtAl2024, MolinaroAndCollins2023GoalCentric}. This flexibility, however, often comes at the cost of efficiency. In this study, we introduced a computational framework that formalizes such trade-offs. Our model posits that goal-dependent learning is initially supported by a fast, but costly working memory (WM) system that maintains goal representations to evaluate outcomes and inform the reinforcement learning (RL) process \cite{McDougleEtAl2022}. While this allows for one-shot value attribution, it inevitably takes away from the limited capacity of WM, at the cost of either less robust or less precise task-relevant representations, which ultimately impair performance \cite{Oberauer2019}. 
With consistent experience, however, learners can create a ``compressed'' reward function -- a simplified representation containing only the essential features needed to identify a goal -- which can then be transferred to long-term memory, enabling the reallocation of WM resources and boosting learning efficiency.

Through a series of experiments and computational modeling, we found support for this framework. 
Experiment 1 demonstrated that goal-dependent learning is parametrically impaired by the number of possible goals: as this number increased within the expected range of WM capacity, learning accuracy decreased. This finding aligns with classic research showing the capacity-limited nature of WM and its impact on concurrent tasks \cite{Baddeley1998, LuckAndVogel1997}, and shows that participants seem to maintain the goal space in WM, despite not needing to do so, since goals were instructed on each trial. This active maintenance is a prerequisite for leveraging executive functions to build a goal-dependent reward function. 
By keeping the number of goal outcomes constant but varying the complexity of the rule defining a goal, we showed in Experiment 2 that learning significantly improves when the goal space is ``compressible''. This suggests that performance improvements are not merely a function of repeated exposure to the same goals, but rather depend on the ability to extract a simple, generalizable reward function -- especially when the goals themselves are too complex to be fully stored in WM. Several mechanisms might support reward function compression, including attention-based feature selection \cite{NivEtAl2015, LeongEtAl2017}. While this specific study allowed for associations between success and nameable rules, similar processes might apply to differently learned reward functions, which may or may not require compression.
In addition, the observed correlation between faster reward collection and better learning performance showed that more efficient reward processing interferes less with the resource-demanding process of task pursuit, leading to more efficient learning.

Our findings also rule out competing explanations.
In their initial study, \cite{McDougleEtAl2022} proposed that the difference in learning rate between Points and Goals could be attributed to occasional lapses occurring in Goals trials, where participants failed to encode or recognize the goal image and subsequently update the probability of repeating the chosen response in future iterations. Excluding this possibility, Supplementary Experiments 3-4 showed excellent goal outcome recognition in spite of performance differences.
In principle, dynamic performance improvements over the task horizon could be explained by cached estimates for the value of an initially neutral outcome (the goal image) that becomes more predictive of reward over time (reminiscent of temporal difference learning; see \cite{GershmanEtAl2014, Sutton1991} for similar principles). However, Supplementary Experiments 5 and 6 showed that manipulating the number of times a goal image is experienced does not, by itself, improve learning -- in line with our proposal that reward function compression is necessary for performance gains. Nonetheless, we note that repetition likely fosters long-term reward functions with small enough goal spaces. 
Another possibility is that goal-dependent reward attribution is done in a completely cost-free, top-down manner that does not take the entire goal spaces into account. By illustrating that performance drops when goal/nongoal labels for the same outcome image are inconsistent with each other, Supplementary Experiment 6 excludes this possibility. 
Finally, computational modeling in Experiments 1-2 showed that the performance bottleneck is not at the decision-making step, but rather at the outcome evaluation and integration stages. 

Our findings extend a growing body of literature highlighting the interplay between executive functions and RL \cite{CollinsAndFrank2012, Collins2018, RmusEtAl2021, YooAndCollins2022}. WM in particular is required to support several aspects of RL, including binding relevant elements of the task at hand, retrieving recent episodes, and even replacing RL entirely \cite{Collins2018, HamEtAl2025, RmusEtAl2021, YooAndCollins2022, YooEtAl2023}. Here, we focus on the less explored aspect of reward function construction and utilization. We build on prior studies showing that learning from novel, abstract goals is possible but effortful \cite{BraemEtAl2024, McDougleEtAl2022, MolinaroAndCollins2023Hacks}, and propose a specific mechanism -- reward function compression -- to explain how this initial cost can be overcome. 

Our suggested process of reward function compression and transfer to long-term memory is consistent with long-standing theories, which established that consistent practice allows cognitive processes to shift from controlled, effortful search to automatic, effortless retrieval from memory \cite{CarlisleEtAl2011, ShiffrinAndSchneider1977, WoodmanEtAl2013}. For this to occur, the information maintained in WM must be consistent \cite{Logan1988}. 
A similar dynamic occurs in action-effect learning, where outcomes produced by responses gradually become incorporated into action representations within a few trials \cite{WolfenstellerAndRuge2011}, but only under restricted conditions \cite{JanczykEtAl2024}.
In our paradigm, it is only when the goal space is stable and simple enough to be held in WM (as in our single goal pair condition of Experiment 1 or the ``Compressible'' condition of Experiment 2) that a consistent mapping from outcomes to value can be learned and automatized. However, a formal model of reward functions' long-term memory transfer remains to be established.

In real life, a major challenge of pursuing arbitrary goals is that they exist in a high-dimensional state space where most features are uninformative \cite{Shenhav2024}. Our work suggests that a key mechanism for efficient learning in such environments is the ability to compress information about rewarding outcomes, ignoring irrelevant features to facilitate generalization and valuation. This resonates with philosophical concepts like ``value capture'' where individuals gravitate towards simple, quantifiable metrics, sometimes at the expense of the richer, more complex values they are meant to represent \cite{Nguyen2024}. Our work suggests ``compressed'' reward functions are sought because they help reduce cognitive demands and ensure learning efficiency -- albeit towards a different behavior than initially intended.

Our work also opens several further questions. The ``goals'' in our experiments were instructed and highly simplified. Future research should investigate whether the principles of reward function compression apply in more naturalistic settings, especially with self-imposed goals. 
Furthermore, the process by which people create compressed reward functions from raw example outcomes -- which we did not model in detail -- remains to be elucidated. Follow-up work in this direction might benefit from existing literature on function learning. For instance, reward functions may be created based on learned rules, judgments of similarity, or a mixture of the two \cite{Carroll1963, DeLoshEtAl1997, LucasEtAl2015, McDanielAndBusemeyer2005}.
It would also be interesting to test whether lossy reward function compression leads to systematic errors in value generalization to newly experienced outcomes. More generally, future work should propose and test a model of how reward function compression occurs dynamically on a trial-by-trial basis, and how it is implemented neurally.
Moreover, the present study focused on goal-dependent values as a third kind of reinforcer, but goals might also impact primary or secondary rewards \cite{JuechemsAndSummerfield2019, KeramatiAndGutkin2014, PanEtAl2025}. Systematic differences in reward function compression between deterministic tasks (such as the ones presented here) and probabilistic goal-dependent RL tasks \cite{McDougleEtAl2022} also remain to be tested. 
Another avenue for future research would thus be to test the directionality of the association between the performance differences predicted by the reward function compression hypothesis and reward collection RTs, since we can only claim correlation, not causation. For example, faster reward collection RTs could be due to learning improvements leading participants to expect desirable outcomes more reliably.
Finally, our sample was relatively uniform, consisting of university undergraduates. While our exclusion rates were typical for online studies of similar difficulty \cite{LiAndCollins2025, PanEtAl2025, VaidyaEtAl2021}, we cannot make conclusions about the applicability of our results to less attentive participants.

In conclusion, of our work provides a formal model and empirical evidence for how humans manage the trade-off between flexibility and efficiency in goal-directed learning. We demonstrate that the ability to learn from novel goal-conditioned outcomes is constrained by WM capacity and facilitated by the creation of compressed reward functions that can be transferred to long-term memory. By offering new insights into the cognitive science of motivation, our study provides a potential avenue for developing behavioral and educational strategies to help people achieve their ambitions.

\section*{Methods}
\subsection*{Data collection}
All participants were recruited through the University of California, Berkeley's pool of participants (RPP). Upon providing informed consent, participants completed the experiment online from their own devices. The University of California, Berkeley Institutional Review Board approved the experimental procedure (protocol \#2024-04-17357).

To reduce spurious effects due to inattentive participants, exclusion criteria were devised based on visually identified elbow points \cite{XiaEtAl2021}. Participants were excluded if they: 1) failed to respond on more than 25 trials in total, 2) missed more than 10 trials consecutively, 3) selected the same response more than 15 times in a row, 4) failed to accurately collect rewards in Points trials, as evidenced by either a \textit{d'} below or equal to 1, or more than 20 reward collection errors, 5) did not understand the instructions for reward collection in Goals trials, as evidenced by a reward collection\textit{d'} below or equal to 0, or more than 50 reward collection errors in the Goals condition, 6) failed to collect rewards entirely in one or more blocks, 7) had an average response time of less than 300 ms in either the Points or the Goals condition (or both). Participants under one or more such categories were excluded from further analysis. In each new experiment, we aimed to gather data from about 50 participants after applying exclusion criteria. 

\subsection*{Data analysis}
Unless otherwise specified, statistical tests were run using the \texttt{scipy} package \cite{VirtanenEtAl2020}.
We used double-sided t-tests to compare average learning performance across pairs of outcome conditions. 
When comparing average scores across different experiments with unequal variances, we applied Welch's correction. 
For null t-test results, we derived Bayes factors via the \texttt{pingouin} package \cite{Vallat2018}, and interpreted them following \cite{LeeAndWagenmakers2014}.
All reported correlations are computed as Spearman's $\rho$.
Results for mixed-effects models (MEMs) were estimated via the \texttt{pymer4} package \cite{Eshin2018}. All continuous variables entered in MEMs were first standardized. When block numbers were added to the MEM, they refer to the block number within the corresponding Points or Goals condition. Whenever stimulus iterations were added to the MEM, they were entered as raw values, but we obtained similar results when first taking the log. Categorical variables were first turned into continuous, with the following coding scheme. When comparing between Points and Goals, block conditions were coded as 0.5 and -0.5, respectively. When comparing different conditions across Goals blocks, we used 0, -0.5, -1, and -1.5 for one, two, three, and four-pair blocks in Experiment 1 and 0.5 and -0.5 for the Compressible and Incompressible blocks in Experiment 2. When comparing different phases within Experiment 2, we used -0.5 for the first set of blocks and 0.5 for the second.

\subsection*{Computational modeling}


All models shared the base mechanisms of a reinforcement learning (RL) agent that tracks $Q$ values for each stimulus and response, yielding an $N_{s}\times N_{a}$ table where $N_{s} = 6$ is the number of stimuli and $N_{a} = 3$ is the number of possible responses in each block. Values are initialized at $Q_{0} = 1/N_{a}$ at the beginning of each block, i.e., with uniform probability. 
Upon receiving feedback $r$ on trial $t$, the value of the chosen response ($a_{c}$) to the presented stimulus ($s$) is updated via the delta rule \cite{RescorlaAndWagner1972}:
\begin{equation} \label{eq:delta_rule}
\delta_{s,a_{c},t} = r_{s,a_{c},t}-Q_{t}(s,a_{c})
\end{equation} 
\begin{equation}
Q_{t+1}(s,a_{c})=Q_{t}(s,a_{c})+\alpha\cdot\delta_{s,a_{c},t}
\end{equation}
where $\alpha$ is the learning rate. A multiplier $\zeta$ was applied to the learning rate when outcomes were negative \cite{LefebvreEtAl2017}.
$Q$ values were converted to a policy via the softmax function \cite{Luce1959} with inverse temperature $\beta$ and the noise parameter $\epsilon$ to capture random errors in the participants' choices:
\begin{equation} \label{eq:softmax}
P_{t}(a_{c}|s) = (1-\epsilon)\frac{\exp(\beta\cdot Q_{t}(s,a_{c}))}{\sum_i \exp(\beta\cdot Q_{t}(s,a_{i}))} + \epsilon \frac{1}{N_{a}}
\end{equation} 
Because it trades off with other parameters (e.g., the learning rate -- potentially undermining parameter recoverability), $\beta = 25$ was a fixed parameter in all models, as is common practice with similar models and tasks \cite{CollinsAndFrank2018, WestbrookEtAl2024}. Note that decision noise is captured by the free parameter $\epsilon$ mentioned above. 
On each trial, a forgetting rate $\phi$ brought $Q$ values closer to $1/N_{a}$ \cite{McDougleEtAl2022}:
\begin{equation}
Q_{t} = Q_{t} + \phi (Q_{0}-Q{t} )
\end{equation}

After a decision was made and $Q$ values were updated, a ``stickiness'' value $\eta \in [-1,1]$ was added to $Q_{t}(s,a_c)$, where $a_c$ is the recently chosen action \cite{SugawaraAndKtahira2021}. This yielded a combined value table $W$, which entered the softmax (Equation \ref{eq:softmax}) on the next trial and made the model biased towards (or against) the previous response in a stimulus- and outcome-insensitive fashion. Unchosen actions were left unmodified. Hence:
\begin{equation}
W_{t+1,a_i} = 
\begin{cases}
Q_{t,a_i} + \eta \text{ if }  a_i = a_c  \\
Q_{t,a_i} \text{ if }  a_i \neq a_c 
\end{cases}
\end{equation}
Stickiness thus captures biases in participants' choice selection which drove them to either replicate or avoid the most recently chosen action irrespective of the outcome or the stimulus being presented. 

Models also kept track of chosen stimulus-response pairs in a stimulus-dependent, but outcome-independent fashion \cite{Collins2026}. This was implemented through a separate table of values, called $C$, for each stimulus and response. After being initialized at $C = 0$, the value of the chosen stimulus-action pair was updated via the delta rule, but ignoring the outcome:
\begin{equation}
C_{t+1}(s,a_{c})=C_{t}(s,a_{c})+\tau\cdot(1-C_{t}(s,a_{c}))
\end{equation}
where $\tau$ is the learning rate. 
$C$ values were mixed with $Q$ values as determined by the value of parameter $\kappa$ before entering the softmax (Equation \ref{eq:softmax}): 
\begin{equation} \label{eq:out_blind}
    M_{t} = (1-\kappa )Q_{t} + \kappa C_{t}
\end{equation}
In models with stickiness, $Q_{t}$ in Equation \ref{eq:out_blind} was replaced with $W$.

All the parameters listed in the previous sections could optionally vary according to the block's condition (Points or Goals) and within-block sub-conditions (specific manipulations within Goals blocks).
A separate family of models had only one instance of each free parameter, but this value was then scaled by the free parameter $\omega$ and multiplied by the median log reward collection RT (or choice RT, for alternative models) for the corresponding block and condition. This was intended to model how reward compression efficiency impacted the learning process. We use the median log RTs to mitigate the impact of any shared variance between reward collection/choice RT and trial-by-trial noise.

The \texttt{fmincon} function from MATLAB 2022a (via \texttt{createOptimProblem} and Python MATLAB engine) was used to obtain the best-fitting parameters for all computational models. The function was used to optimize the log-likelihood of the data under the specified model architecture \cite{WilsonAndCollins2019}. All free parameters were bounded between 0 and 1, except for $\eta$ (stickiness parameter), which could take any value between -1 and 1 (since persistence biases are often found, but our deterministic task design could encourage switch behavior)
and $\omega$, which was constrained between -10 and 10.

Models were evaluated in terms of both predictive and generative validity \cite{PalminteriEtAl2017}. 
We used the Akaike information criterion (AIC) to assess each model's relative ability to predict the observed data.
For differences in model AICs, we report the Wilcoxon signed-rank test statistic.
To assess each model's ability to reproduce the effects of interest, we simulate its behavior with the best-fit 20 times per participant, amounting to roughly 1000 total iterations per model. 

We verified that models were identifiable in all key comparisons (Supplementary Fig. 10), and that the winning model's best-fitting parameters (Supplementary Fig. 11) were recoverable (Supplementary Fig. 12).
\label{sec:methods}



\bibliographystyle{naturemag}
\bibliography{references}

\section*{Acknowledgments}
This work was funded by NSF Grants 2020844 and 2336466 awarded to A.G.E.C. We are grateful to Samuel McDougle, Beth Baribault, and Amy Zou for support on previous iterations of this work.

\section*{Author contributions}
G.M. and A.G.E.C. conceptualized the study. G.M. collected, visualized, and analyzed the data. G.M. wrote the manuscript (original draft). G.M. and A.G.E.C. revised and edited the manuscript. A.G.E.C. supervised the study and acquired financial support. 

\section*{Competing interests}
The authors declare no competing interests.
\newpage

\mocktitleheader{Supplementary Information}

\noindent\textbf{Table of Contents}

\startcontents[mocked]
\vspace{-1.25em}
\printcontents[mocked]{l}{1}{\setcounter{tocdepth}{3}}
\vspace{1em}
\hrule 

\startlist[mockedfigures]{lof}

\vspace{1em}
\noindent\textbf{List of Figures}\printlist[mockedfigures]{lof}{l}{}
\vspace{1em}
\hrule

\newpage
\section{Supplementary experiments}
\subsection{Experiment 3} 
Experiment 3 was previously reported in a separate publication \cite{MolinaroAndCollins2023Hacks} and follows the same structure as the other experiments presented in the current article. Unlike Experiments 1-2, but similar to \cite{McDougleEtAl2022}, goal and non-goal images changed on every trial (Figure \ref{fig:behav_results_3_to_6}a). The six blocks of the experiments were equally split among the Points and Goals conditions.
Here, we extend previous analyses on the same data set to show that, when reward function compression is prevented (by having goal images change on each trial), learning performance remains inefficient compared to learning from established rewards. Moreover, we show that lapses in goal encoding or recognition that prevent subsequent learning are rare and unlikely to cause the observed differences in performance. To do so, we leverage the ``reward collection'' stage of our task, which required participants to press a dedicated key upon receiving positive feedback.

One hundred and twenty-five participants completed Experiment 3. Thirty-five participants failed to meet baseline requirements and were excluded from further analysis. Therefore, data from 90 participants (70\% female, ages 18-30, M = 20.33 $\pm$ 0.21) were analyzed. 

As previously observed in a similar task \cite{McDougleEtAl2022}, we found that performance in Goals blocks was above chance (i.e., 33\%; M = 0.67 $\pm$ 0.02, $\mu$ = 0.33; t(89) = 21.94, p $<$ 0.001), meaning participants successfully learned the correct stimulus-response mapping from abstract, novel, neutral images. 
At the same time, participants' performance in Points blocks (M = 0.8 $\pm$ 0.01) was significantly higher than in Goals blocks (t(89) = 12.65, p $<$ 0.001; Figure \ref{fig:behav_results_3_to_6}b). 
When feedback was delivered in the form of numeric points, learning accuracy proceeded faster and reached a higher asymptote (Figure \ref{fig:behav_results_3_to_6}c). 
To compare learning speed in Goals vs. Points blocks, we fit a linear mixed-effects model (MEM) with random intercepts for participant number, predicting average learning performance from condition, stimulus iteration, and within-condition block number. In addition to main effects of condition ($\beta$ = 0.125 $\pm$ 0.01, 95\% CI = [0.12, 0.13], p $<$ 0.001), stimulus iteration ($\beta$ = 0.14 $\pm$ 0.002, 95\% CI = [0.13, 0.14], p $<$ 0.001), and block number ($\beta$ = 0.04 $\pm$ 0.002, 95\% CI = [0.04, 0.05], p $<$ 0.001), we found a small, but significant interaction between block condition and stimulus iteration ($\beta$ = 0.01 $\pm$ 0.01, 95\% CI = [0.003, 0.02], p = 0.009), with performance improving faster for Points than Goals.
Consistent with this, performance was higher for Points (M = 0.93 $\pm$ 0.01) than Goals (M = 0.82 $\pm$ 0.02; t(89) = 7.74, p $<$ 0.001) even in the last two stimulus iterations of each block.
In addition, we found a significant interaction between condition and block number ($\beta$ = -0.06 $\pm$ 0.01, 95\% CI = [-0.07, -0.06], p $<$ 0.001), with performance benefiting from additional blocks more in the Goals condition than in the Points condition. 
Nonetheless, differences in performance between Points (M = 0.81 $\pm$ 0.01) and Goals (M = 0.75 $\pm$ 0.02) remained significant even in the last block of each condition (t(89) = 4.33, p $<$ 0.001).


Overall, reward collection accuracy (measured as $d^\prime$ -- the difference between the Z-scored hit and false alarm rates) was well above chance (i.e., 0; M = 3.62 $\pm$ 0.06; t(89) = 59.75, p $<$ 0.001), confirming participants understood the reward collection aspect of the task.
While there was a statistically significant difference in reward collection accuracy between Points (M = 3.99 $\pm$ 0.07) and Goals (M = 3.4 $\pm$ 0.07; t(89) = 7.6, p $<$ 0.001), 
even the latter was close to ceiling, with 95\% of Goals reward collection responses being correct (vs. 98\% for Points; Figure \ref{fig:behav_results_3_to_6}d). 
If lapses in goal encoding or recognition were the sole culprit for learning inefficiencies in Goals trials, we should observe a correlation between condition-based differences in reward collection accuracy and learning performance. 
However, we found no such correlation ($\rho$(89) = 0.18, p = 0.086). 
In previous work \cite{MolinaroAndCollins2023Hacks}, we also found that a computational model with subjective rewards based on reward collection decisions (i.e., +1 when participants ``collected'' and +0 when they did not) did not improve performance relative to a baseline model with objective outcome-based feedback. 
Therefore, occasional errors in positive outcome recognition are unlikely to explain the observed difference in learning from goal-contingent outcomes relative to familiar rewards.

In sum, Experiment 3 replicates previous findings showing that humans can flexibly imbue neutral outcomes with value, hence leveraging them as learning signals, but that such flexibility comes at a cost \cite{McDougleEtAl2022}. 
Although plausible, we did not find evidence for a strong contribution of lapses in goal encoding to learning impairments. 

\subsection{Experiment 4} 
Experiment 4 was also previously reported \cite{MolinaroAndCollins2023Hacks}. In this version of the experiment, the same two fractals were used as goal and nongoal images throughout the experiment  (Figure \ref{fig:behav_results_3_to_6}e). Pairs of fractals were created to maximize visual distinctiveness and randomized across participants. Goal and nongoal images were assigned during the instructions, as well as the start of each trial in Goals blocks. As in Experiment 3, participants completed three Points and three Goals blocks, pseudo-randomly interleaved.
Keeping a single goal outcome for the whole experiment enabled us to test differences in goal-dependent and point-dependent learning when reward function compression would be easiest to perform. 
Indeed, in this manipulation, participants only needed to identify a key, distinguishing feature of the goal image to evaluate the outcome (e.g., its dominant color, pattern, etc.). Such information could eventually be transferred into a reward function and stored in long-term memory, making more executive resources available for learning. In addition to performing the main task, participants estimated their liking of goal and nongoal images on Likert scales from 1 (``Not at all'') to 5 (``Very much'') at three distinct points of the experiment: 1) before any instructions were given on the upcoming task, 2) after participants were informed of which fractal would be used as goal vs. nongoal feedback throughout the experiment, and 3) at the end of the experiment.

One hundred and twenty-one participants completed Experiment 4. The final data set after exclusion comprised 93 participants (73\% female, ages 18-27, M = 20.09 $\pm$ 0.19).

Using a single pair of fractals as goal/nongoal images led to a substantial increase in Goals in learning compared to Experiment 3, with the difference between Points and Goals accuracy dropping significantly (from M = 0.13 $\pm$ 0.01 to M = 0.03 $\pm$ 0.01; t(181) = 7.24, p $<$ 0.001). This drop was specifically driven by the Goals condition (Experiment 3 M = 0.67 $\pm$ 0.02, M = 0.77 $\pm$ 0.01; t(181) = -4.99, p $<$ 0.001), whereas Points performance did not significantly differ between the two experiments (M = 0.8 $\pm$ 0.01, M = 0.8 $\pm$ 0.01; t(181) = 0.01, p = 0.994, BF10 = 0.16 suggesting moderate evidence for the null hypothesis). 

Although smaller in magnitude compared to Experiment 3, the average difference between Points (M = 0.8 $\pm$ 0.01) and Goals learning accuracy (M = 0.77 $\pm$ 0.01) remained significant (t(92) = 3.56, p $<$ 0.001; Figure \ref{fig:behav_results_3_to_6}f). 
By contrast, reward collection accuracies (measured as $d^\prime$) did not differ between the two conditions (Points M = 3.96 $\pm$ 0.05, Goals M = 3.93 $\pm$ 0.05, t(92) = 0.43, p = 0.667, BF10 = 0.13 moderate evidence for the null hypothesis; Figure \ref{fig:behav_results_3_to_6}h) -- a result which further discredits the hypothesis that lapses in outcome encoding or recognition are the main cause of learning impairments in the Goals condition.
Unlike in Experiment 3, learning did not proceed at different rates within Points compared to Goals blocks ($\beta$ = -0.01 $\pm$ 0.004, 95\% CI = [-0.02, 0.003], p = 0.007]; Figure \ref{fig:behav_results_3_to_6}g). 
On the contrary, while performance in Goals trials was initially worse than in Points trials, it eventually reached similar levels, such that choice accuracy in the two conditions was no longer significantly different in the last two stimulus iterations within a block (Points M = 0.93 $\pm$ 0.01, Goals M = 0.92 $\pm$ 0.01; t(92) = 1.65, p = 0.103, BF10 = 0.42, anecdotal evidence for the null hypothesis). We again found a significant interaction between condition and block number, suggesting stronger meta-learning in Goals than Points trials ($\beta$ = -0.16 $\pm$ 0.004, 95\% CI = [-0.03, -0.01, p $<$ 0.001]). 
Unlike in Experiment 3, we found no significant performance differences between the two conditions in the last block (Points M 0.82 $\pm$ 0.01, Goals M = 0.81 $\pm$ 0.01; t(92) = 1.01, p = 0.317, BF10 = 0.19, moderate evidence for the null hypothesis). 
All MEM effects of condition ($\beta$ = 0.03 $\pm$ 0.004, 95\% CI = [0.02, 0.04], p $<$ 0.001), stimulus iteration ($\beta$ = 0.14 $\pm$ 0.002, 95\% CI = [0.13, 0.15], p $<$ 0.001), and block number ($\beta$ = 0.04 $\pm$ 0.002, 95\% CI = [0.04, 0.05], p $<$ 0.001) were consistent with our findings in Experiment 1. 
Together, these results suggest that learning from a single outcome, compared to several different instances, benefits goal-conditioned learning. 

Supporting the idea that iteratively assigning value to the same goal outcome increases its appetitive value, differences in the subjective liking of goal images, compared to nongoal images, significantly increased from the beginning (M = 0.05 $\pm$ 0.15) to the end of the task (M = 0.65 $\pm$ 0.18; Z(92) = 219.5, p $<$ 0.001), when they were significantly above 0 (Z(92) = 954.5, p = 0.001).

In sum, goal-dependent learning greatly improved when abstract goal outcomes were fixed, compared to when they were instructed on each trial. This result highlights that imbuing neutral stimuli with value is costly, potentially withdrawing resources from the learning process. With more experience, performance in Goals was not significantly different from that of Points trials. Subjective valuations of goal, relative to nongoal images, also increased throughout the task. 
These findings are in line with the idea that, given a consistent set of goal features, people can compress the information required to recognize desirable outcomes and spare WM resources to support learning. 

\subsection{Experiment 5}
In principle, the fact that by the end of Experiment 4, participants could perform similarly in the Goals and Points condition could be explained through an iterative process, which facilitates learning from abstract goals with every additional experience. Such a process could, in theory, rely on cached estimates for the value of an initially neutral outcome that becomes more predictive of reward over time (see \cite{GershmanEtAl2014, Sutton1991} for similar principles). 
Experiment 5 directly tested this hypothesis by manipulating the number of times a goal image was repeated for different stimuli in a block. Unbeknownst to participants, the six stimuli in each Goals block were split into three groups, where the same goal/nongoal image pairs were used as feedback 75\%, 50\%, or 25\% of the time, while the rest of the trials involved a new pair of images (Figure \ref{fig:behav_results_3_to_6}j). 
Repeated fractal pairs were unique to each group and shared across stimuli of the same group across blocks, such that participants could not prepare responses during the initial presentation of goal/nongoal outcomes. We additionally ensured that the correct stimulus-response association was distinct for stimuli of the same group. Repeated fractal pairs were re-utilized for the same group assignment across blocks, whereas novel fractal pairs were unique to each trial in which they were employed. Outcome images and stimulus assignments for each block, condition, and group were pseudo-randomized. 

Sixty-seven participants completed Experiment 5, 16 of whom did not meet our inclusion criteria. Data from 51 subjects were analyzed (80\% female, ages 18-25, M = 20.68 $\pm$ 1.30).

If goal-contingent value assignment benefited from merely attempting the same goal multiple times, we should expect a significant difference across stimulus groups in the Goals blocks of Experiment 5. 
To test this, we ran a linear MEM predicting average learning performance in Goals blocks from stimulus groups, with random intercepts for participant number. Contrary to the repetitions hypothesis, we found no main effect of stimulus group ($\beta$ = 0.02 $\pm$ 0.02, 95\% CI = [-0.02, 0.06], p = 0.355; Figure \ref{fig:behav_results_3_to_6}k-l). 
Specifically, although we found a significant difference between the 75\% (M = 0.66 $\pm$ 0.02) and the 50\% repetitions groups (M = 0.61 $\pm$ 0.02; t(50) = 2.27, p = 0.027), there were no significant differences between the 75\% and the 25\% (M = 0.64 $\pm$ 0.02, t(50) = 0.97, p = 0.335, BF10 = 0.24, moderate evidence for the null hypothesis), or the 50\% and 25\% repetitions groups (t(50) = 1.68, p = 0.1, BF10 = 0.56, anecdotal evidence for the null hypothesis). 
Similarly, we did not find differences across stimulus groups even when only considering the last two iterations of each block ($\beta$ = -0.001 $\pm$ 0.03, 95\% CI = [-0.06 0.06], p = 0.967). 
The lack of difference in learning across groups could not be attributed to ceiling effects, since the overall difference between Points (M = 0.81 $\pm$ 0.01) and Goals performance (M = 0.64 $\pm$ 0.02) remained statistically significant (t(50) = 10.97, p $<$ 0.001), even when only considering the last two stimulus iterations of each block (Points M = 0.94 $\pm$ 0.01, Goals M = 0.64 $\pm$ 0.02; t(50)=19.92, p $<$ 0.001).
Moreover, overall Goals performance in Experiment 5 was similar to that of Experiment 3 (t(139) = 1.51, p = 0.134, BF10 = 0.52, i.e., anecdotal evidence for the null hypothesis). The same was true when comparing the overall (t(139) = 0.53, BF10 = 0.21, p = 0.597, signifying moderate evidence for the null hypothesis) or late Goals accuracy of Experiment 3 (i.e., in the last two iterations; t(49) = 0.49, p =0.624, BF10 = 0.21, suggesting moderate evidence for the null hypothesis) to that of the 75\% repetitions group of Experiment 5, wherein repetition effects should have been maximal. Learning from standard rewards was also not significantly different between the two experiments (t(139) = -0.4, p = 0.687, BF10 = 0.20 i.e., moderate evidence for the null hypothesis), suggesting the lack of difference in Goals trials was likely not driven by systematic differences between the two samples. 

A similar pattern was found for reward collection $d^\prime$, which did not vary across groups ($\beta$ = 0.13 $\pm$ 0.10, 95\% CI = [-0.06, 0.32]; Figure \ref{fig:behav_results_3_to_6}m
), even when considering the last two iterations alone ($\beta$ = 0.06 $\pm$ 0.08, 95\% CI = [-0.09, 0.22]
).
Overall, this set of results is inconsistent with the idea that, by itself, re-experiencing the same goal more times increases its reinforcing power. 
 
Contrary to the proposition that simply re-experiencing the same outcome as a goal image strengthens its reinforcing properties, any effects of repetitions, if present, were too small to be detected in Experiment 5. Instead, learning performance was similar across stimulus groups in the Goals condition. 

\subsection{Experiment 6}
Experiment 6 was designed to test how learning from goal-dependent rewards might be affected by experiencing conflicting instances of an outcome being positive on some occasions and negative on others. To that end, the six stimuli in each block were divided into three groups of two (Figure \ref{fig:behav_results_3_to_6}n). 
In each block, both stimuli of the ``pure goal repetitions'' group were associated with a single pair of images, each used 12 times per stimulus. This pair of fractals did not differ from block to block. Stimuli of the ``conflicting goal repetitions'' group were also associated with a single pair of images used 12 times per stimulus. However, fractals belonging to the same pair were presented six times (50\% of the time) with a ``Goal'' label and six times (the remaining 50\%) with a ``Nongoal'' label. To tease apart the beneficial effects of outcome repetition from the detrimental effects of valence conflicts due to label inconsistencies, we also applied a ``baseline'' condition to two stimuli. Here, fractal image pairs belonged to a set of two, such that each pair was used six times in total (as opposed to the twelve times of the ``Pure'' goal repetitions group) and always with consistent labels (unlike the ``Conflicting'' goal repetitions group).
Fractal pairs were shared among stimuli of the same group across blocks, and the correct stimulus-response association was distinct for stimuli of the same group.
If participants relied on top-down, single-shot value attribution to the current goal outcome, performance for the Conflicting group of stimuli should resemble that of the Pure and Baseline stimuli.

Seventy-one participants completed Experiment 6, 20 of which did not pass our inclusion criteria. Data from 51 subjects were analyzed (67\% female, ages 19-26, M = 20.76 $\pm$ 1.53).

We fit a linear MEM predicting average learning performance in Goals blocks from stimulus group, with random intercepts for participant number. Contrary to the single-shot value attribution hypothesis, we found a significant main effect of stimulus group on learning performance ($\beta$ = 0.07 $\pm$ 0.02, 95\% CI = [0.03, 0.17], p $<$ 0.001; Figure \ref{fig:behav_results_3_to_6}o-p). 
More specifically, there was a significant difference in performance between the Pure (M = 0.62 $\pm$ 0.02) and Conflicting (M = 0.55 $\pm$ 0.02) Goals stimulus groups (t(50) = 3.34, p = 0.002). There was also a significant difference between the Baseline (M = 0.61 $\pm$ 0.02) and Conflicting stimulus groups (t(50) = -2.99, p = 0.004). By contrast, learning did not differ between Pure and Baseline stimulus groups (t(50) = 0.64, p = 0.522, BF10 = 0.19, moderate evidence for the null hypothesis). 
Similarly, we found a significant overall effect of stimulus group on performance in the last two stimulus iterations of each block ($\beta$ = 0.08 $\pm$ 0.03, 95\% CI = [0.02, 0.14], p = 0.007). 
On such trials, performance was higher in the Pure (M = 0.76 $\pm$ 0.03) relative to the Conflicting (M = 0.55 $\pm$ 0.02) Goals stimulus groups (t(50) = 2.63, p = 0.011), with no statistical difference between the Pure and Baseline (M = 0.71 $\pm$ 0.04) outcome conditions (t(50) = 1.61, p = 0.113, BF10 = 0.51, anecdotal evidence for the null hypothesis). However, differences between the Baseline and Conflicting stimulus groups did not reach statistical significance (t(50) = -1.14, p = 0.262, BF10 = 0.28, moderate evidence for the null hypothesis).

Reward collection accuracy, measured as $d^\prime$, also varied by group ($\beta$ = 0.71, $\pm$ 0.11, 95\% CI = [0.50, 0.91], p $<$ 0.001; Figure \ref{fig:behav_results_3_to_6}q). Pairwise differences between stimulus groups reflected the results obtained for performance, with significant differences between Pure (M = 3.6 $\pm$ 0.08) and Conflicting (2.89 $\pm$ 0.1; t(50) = 7.14, p $<$ 0.001) and Baseline (3.48 $\pm$ 0.10) and Conflicting stimulus groups (t(50) = -5.49, p $<$ 0.001), but not between Pure and Baseline stimulus groups (t(50) = 1.23, p = 0.225, BF10 = 0.31, moderate evidence for the null hypothesis).  
However, differences in reward collection between Pure and Conflicting stimulus groups were not correlated with the respective difference in learning performance ($\rho$(50) = 0.23, p = 0.101), suggesting outcome mislabeling was not a sufficient explanation for the observed differences among groups.

Confirming our results from Experiments 2 and 5, repeatedly experiencing the same outcome as a positive event did not, by itself, improve learning performance if the goal space remains too vast to compress. By contrast, receiving conflicting information about the value of an outcome across different trials had a negative impact on learning. These results do not fit with the hypothesis that attributing rewarding properties to an otherwise neutral stimulus occurs in a purely top-down, one-shot manner. Instead, the observed pattern of results is consistent with the idea that individuals attempt to integrate goal-dependent information across examples of positive outcomes, and utilize limited cognitive resources doing so. Given that only five unique goal images were used in each block of Experiment 6, participants could, in principle, attempt to identify a ``compressed'' reward function, which would enable them to recognize a desirable outcome as such without having to retrieve its exact perceptual features. However, the ``Conflicting labels'' condition of Experiment 6 would have actively disrupted participants' ability to apply simplified reward functions across goal outcomes, resulting in performance impairments. 

\clearpage
\section{Learning performance comparisons across experiments}
In this section, we report additional learning performance comparisons for Experiments 1-2 compared to relevant control Experiments. 
\subsection{Experiment 1 vs. Experiments 3 and 4}
Performance in the four-pair block of Experiment 1 was similar to that of the first Goals block of Experiment 3 (\cite{MolinaroAndCollins2023Hacks}), where goal/nongoal images changed on every trial (M = 0.57 $\pm$ 0.02; t(142) = -0.60, p = 0.549, BF10 = 0.22, signifying moderate evidence for the null hypothesis). By contrast, performance in the one-pair block was similar to that of the first Goals block in Experiment 4 (\cite{MolinaroAndCollins2023Hacks}, where a single goal/nongoal outcome was employed (M = 0.71 $\pm$ 0.02; t(145) = 0.39, p = 0.697, BF10 = 0.20, signifying moderate evidence for the null hypothesis).  

\subsection{Experiment 2 vs. Experiments 1, 3, and 5}
Because of order effects and differences in the types of goal outcomes employed (fractal images vs. simple color shapes), any direct comparisons between Experiment 2 and other experiments are imperfect. Having taken this cautionary note into account, it is remarkable that learning performance in the second block of the Compressible Goals condition (M = 0.71 $\pm$ 0.02), when putative compressed reward function for outcome to value mapping would have been established, was the same as that of the first Goals block of Experiment 3 (M = 0.71 $\pm$ 0.02; t(142) = -0.22, p = 0.83, BF10 = 0.19, i.e., moderate evidence for the null hypothesis). 
By contrast, learning performance in the last block of the Incompressible Goals conditions (0.61 $\pm$ 0.03) remained lower (t(142) = 3.07, p = 0.003), resembling performance in the four-pair Goals condition of Experiment 1 instead (t(103) = -0.55, p = 0.580, BF10 = 0.24, showing moderate evidence for the null hypothesis). 
Learning in the Compressible goals condition was also initially (i.e., in the first block; 0.56 $\pm$ 0.02) similar to that of learning from four unique pairs of goal-nongoal outcomes in Experiment 1 (t(103) = 0.86, p = 0.394, BF10 = 0.29, suggesting moderate evidence for the null hypothesis). However, once a compressed reward function for this condition was established (i.e., in the first block), Compressible goals performance significantly improved compared to Experiment 1's four-pair Goals condition (t(103) = -3.46, p $<$ 0.001), resembling that of Experiment 5's one-pair Goals condition instead (t(103) = -0.49, p = 0.628, BF10 = 0.23, i.e., moderate evidence for the null hypothesis). 

\clearpage

\section{Supplementary figures}

\begin{figure}[h!]
\begin{center}
\includegraphics[width=0.9\textwidth]{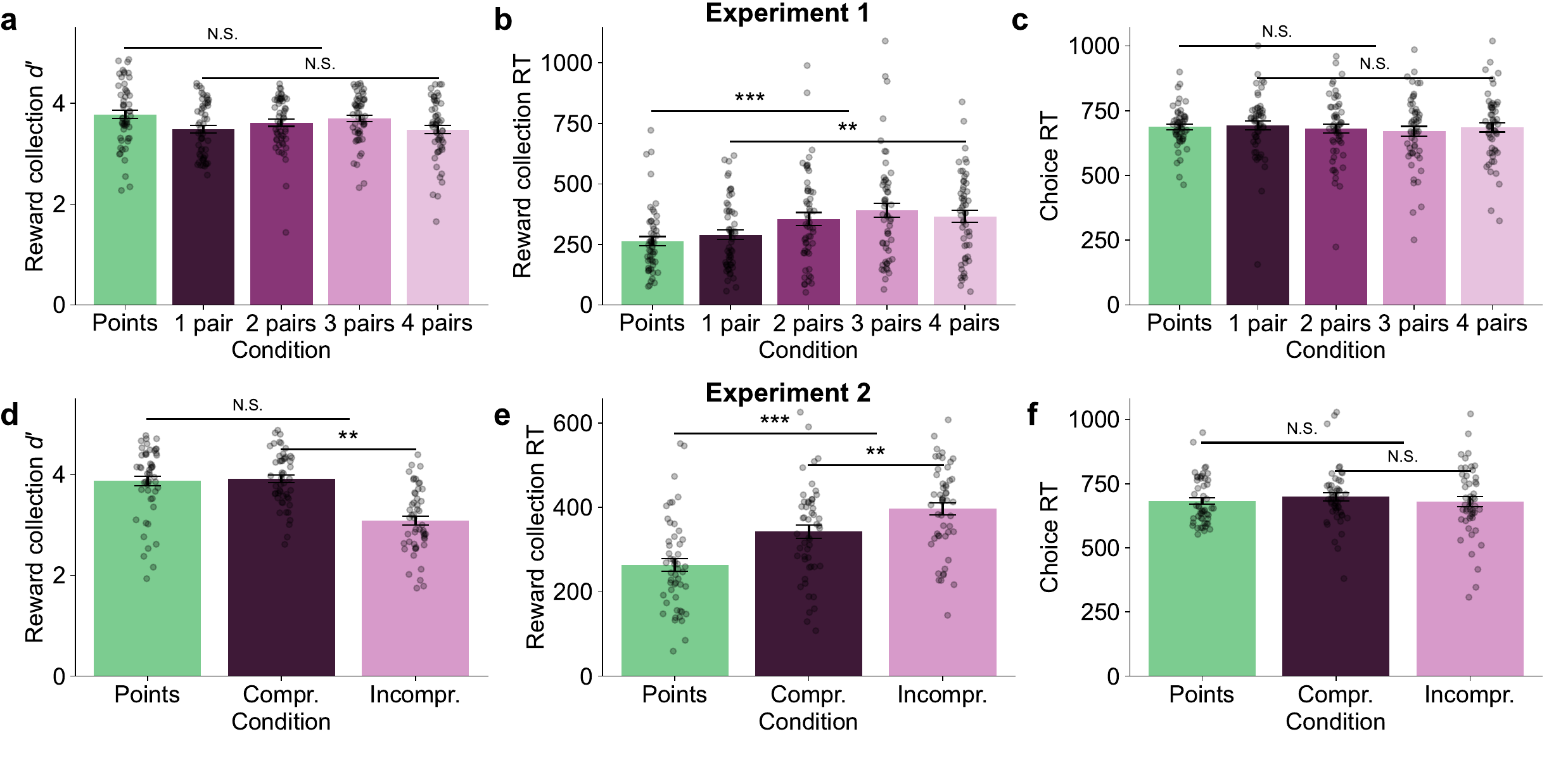}
\end{center}
\caption[Reward collection accuracy and RT for Experiments 1 and 2.]{\textbf{Reward collection accuracy and reaction time (RT; in ms) for Experiments 1 and 2.} \textbf{a, d} Reward collection accuracies across conditions. \textbf{b, e} Reward collection RTs in ms across conditions. \textbf{c, f} Choice RTs across conditions. Bars indicate participants' average behavior, dots show individual participants' scores. Error bars indicate the S.E.M. ** = p $<$ 0.01, *** = p $<$ 0.001}
\label{fig:rc_d_prime_rt_results}
\end{figure}

\clearpage
\begin{figure}[t!]
\begin{center}
\includegraphics[width=0.8\textwidth]{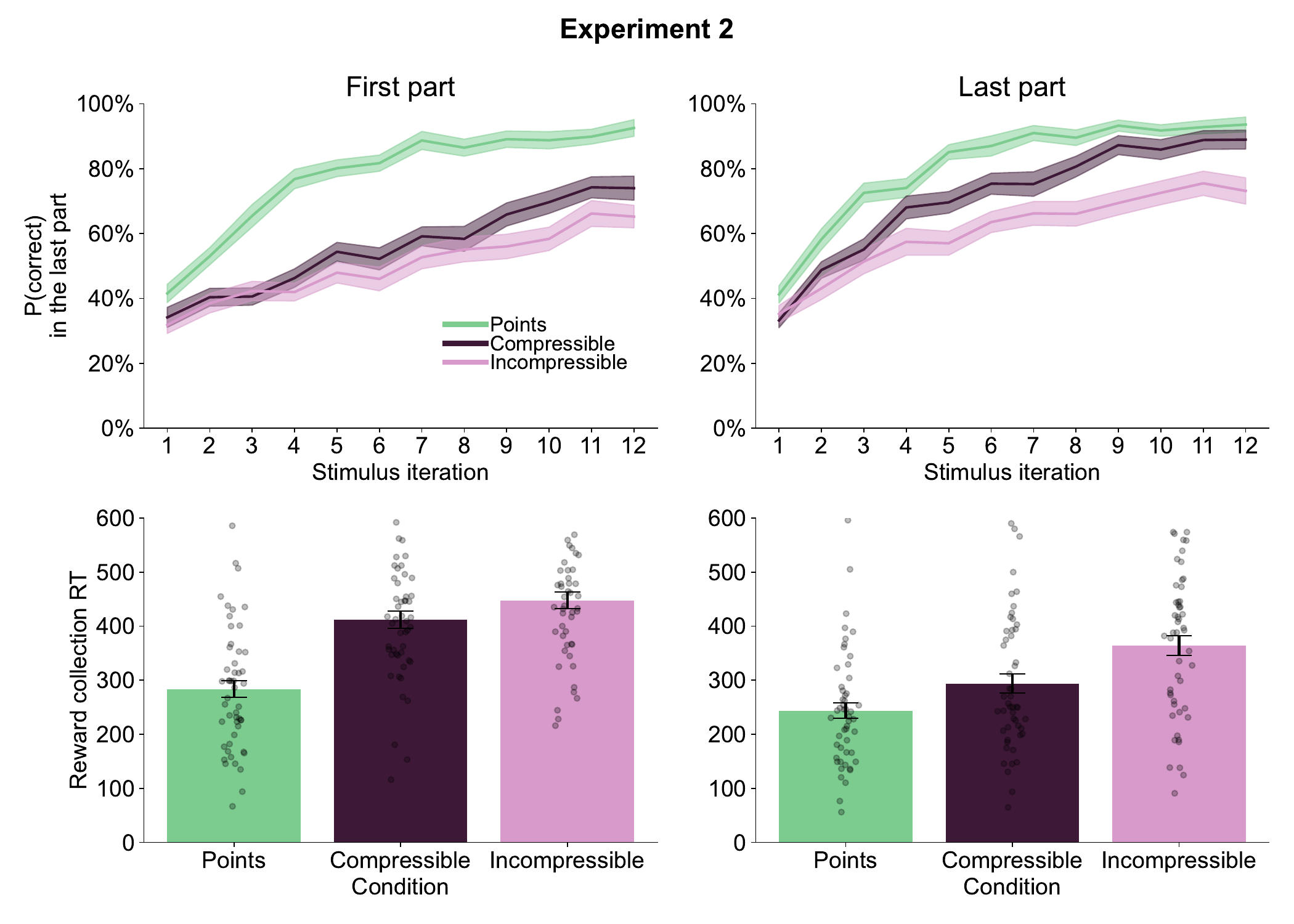}
\end{center}
\caption[Learning curves and reward collection RTs for Experiment 2's first and last set of blocks.]{\textbf{Learning curves and reward collection RTs for Experiment 2's first and last set of blocks.} Bars indicate participants' average behavior, dots show individual participants' scores. Error bars indicate the S.E.M.}
\label{fig:first_vs_last_part_exp_2}
\end{figure}

\clearpage
\begin{figure}[t!]
\begin{center}
\includegraphics[width=1.0\textwidth]{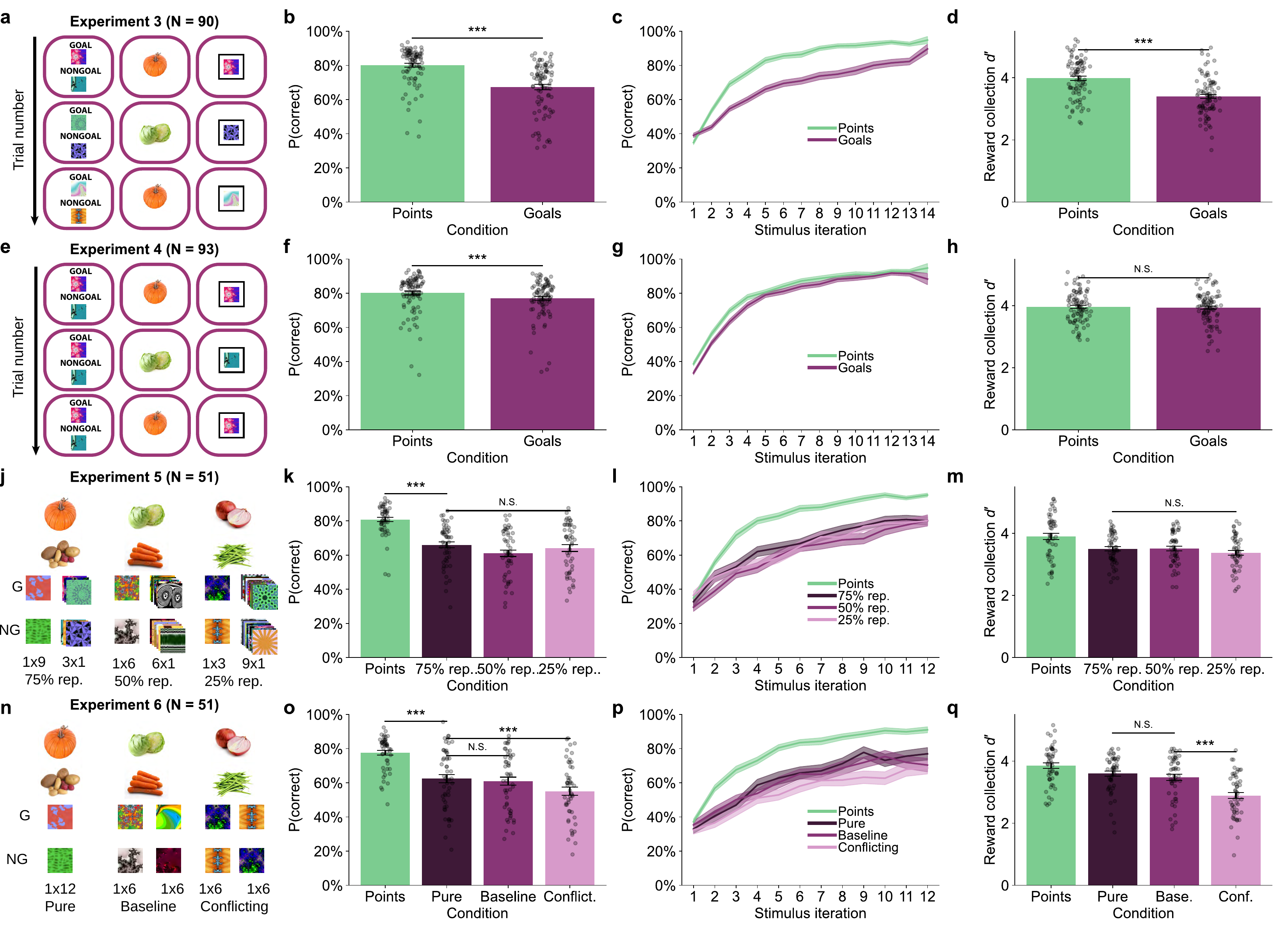}
\end{center}
\caption[Experimental manipulations, learning performance and reward collection accuracy for Experiments 3-6.]{\textbf{Experimental manipulations, learning performance and reward collection accuracy for Experiments 3-6.} \textbf{a, e, j, h} Experimental manipulations for Experiments 3-6. G = Goal image, NG = Nongoal image. \textbf{b, f, k, o} Average learning performance across conditions. \textbf{c, g, l, p} Learning performance over stimulus iterations. \textbf{d, h, m, q} Average reward collection accuracy (\textit{$d^\prime$} across conditions. Bars, dots, and solid lines indicate participants' average behavior. Error bars and shading indicate the S.E.M. *** = p $<$ 0.001}
\label{fig:behav_results_3_to_6}
\end{figure}

\begin{figure}[t!]
\begin{center}
\includegraphics[width=0.7\textwidth]{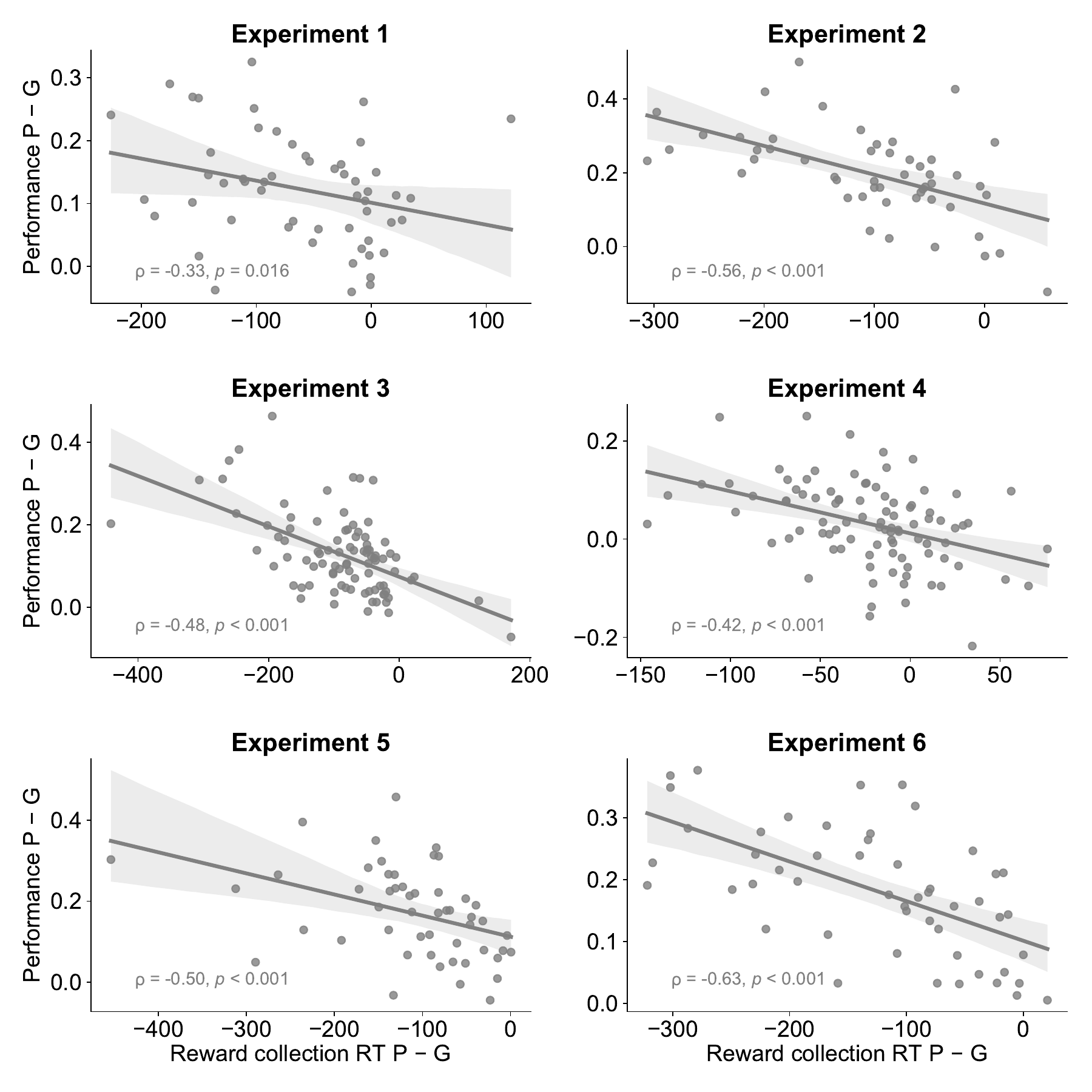}
\end{center}
\caption[Correlations between differences in performance and differences in reward collection RTs across block types in Experiments 1-6.]{\textbf{Correlations between differences in performance and differences in reward collection reaction times (RTs; in ms) across block types (Points vs. Goals) in Experiments 1-6.} Dots represent individual participants. Shading represents 95\% confidence intervals. Spearman's $rho$ correlation coefficients and p-values are shown on each respective plot.}
\label{fig:rc_rt_perf_corrs_by_cond}
\end{figure}

\begin{figure}[t!]
\begin{center}
\includegraphics[width=0.9\textwidth]{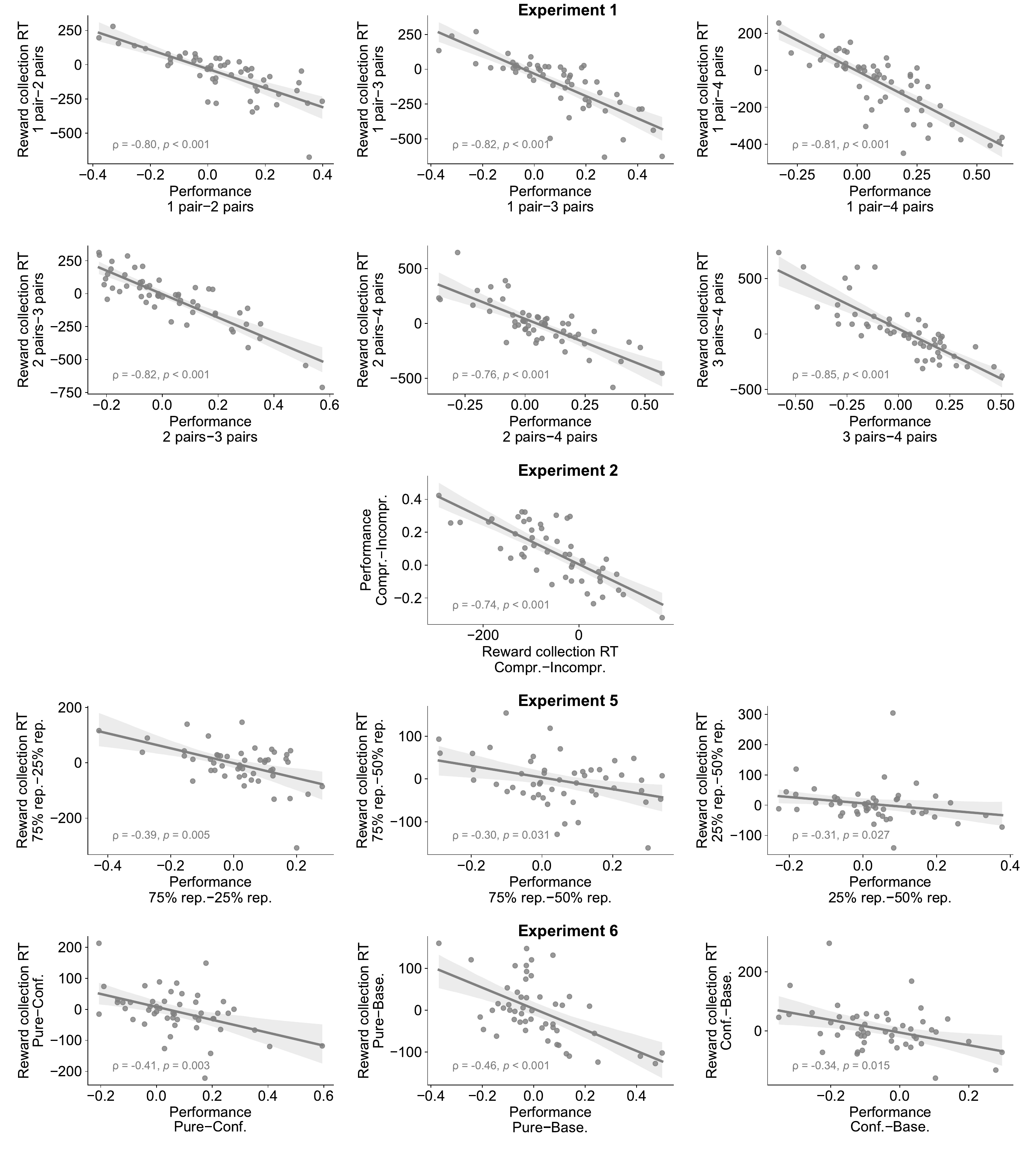}
\end{center}
\caption[Correlations between differences in performance and differences in reward collection RTs across Goals conditions in Experiments 1-2 and 5-6.]{\textbf{Correlations between differences in performance and differences in reward collection reaction times (RTs; in ms) across conditions of Goals blocks in Experiments 1-2 and 5-6.} Dots represent individual participants. Shading represents 95\% confidence intervals. Spearman's $rho$ correlation coefficients and p-values are shown on each respective plot.}
\label{fig:rc_rt_perf_corrs_by_group}
\end{figure}

\begin{figure}[t!]
\begin{center}
\includegraphics[width=0.7\textwidth]{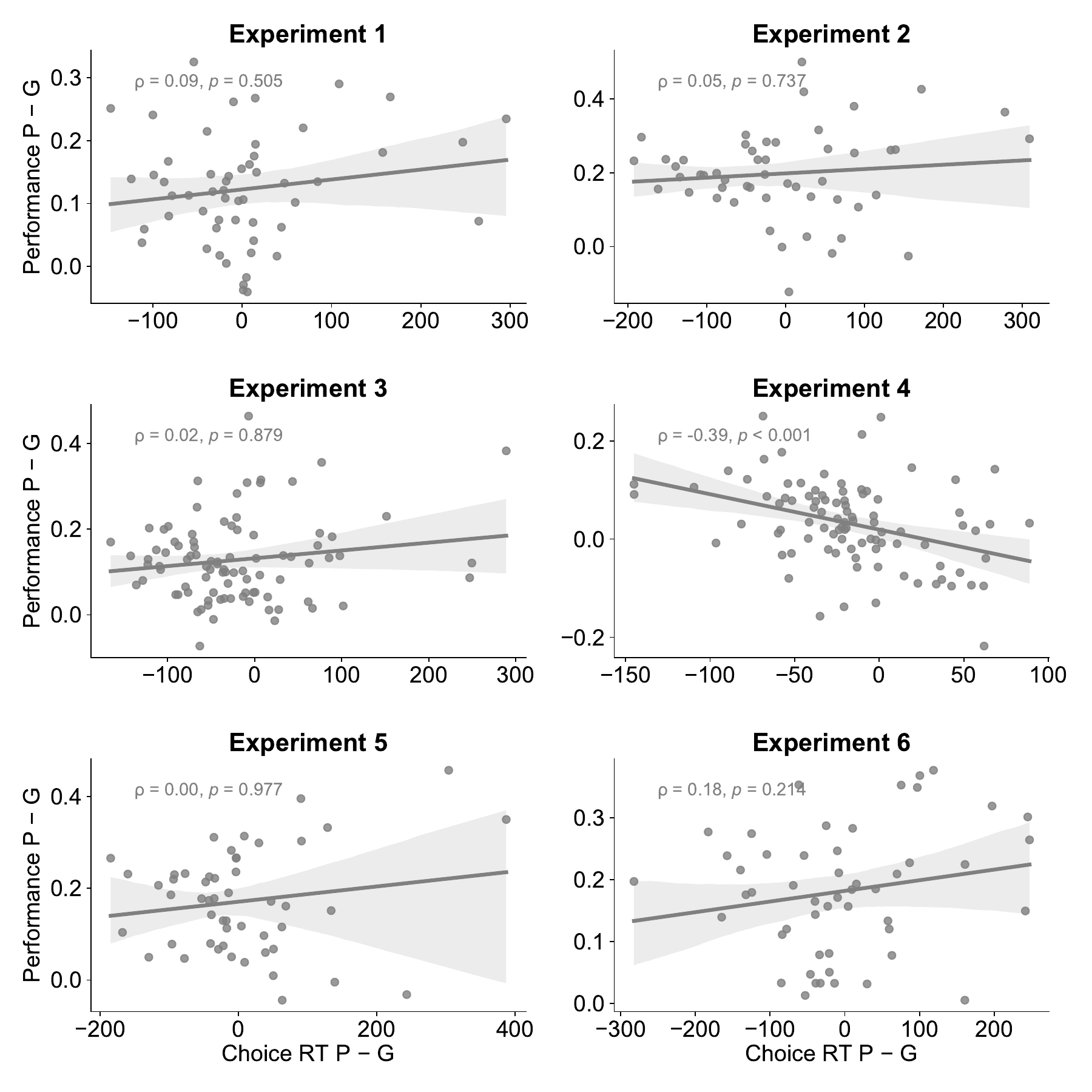}
\end{center}
\caption[Correlations between differences in performance and differences in choice RTs across block types in Experiments 1-6.]{\textbf{Correlations between differences in performance and differences in choice reaction times (RTs; in ms) across block types (Points vs. Goals) in Experiments 1-6. Dots represent individual participants.} Shading represents 95\% confidence intervals. Spearman's $rho$ correlation coefficients and p-values are shown on each respective plot.}
\label{fig:choice_rt_perf_corrs_by_cond}
\end{figure}

\begin{figure}[t!]
\begin{center}
\includegraphics[width=0.9\textwidth]{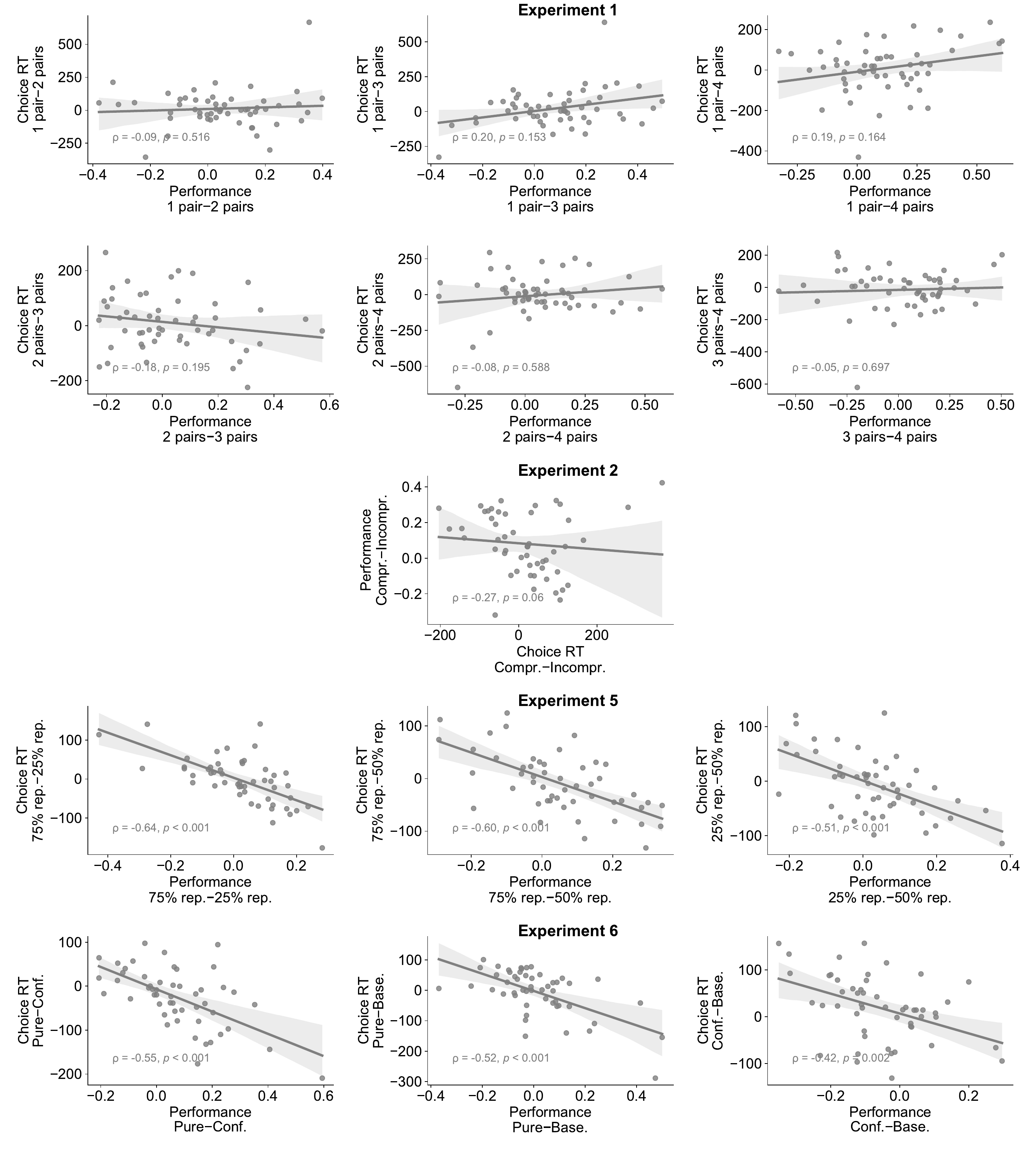}
\end{center}
\caption[Correlations between differences in performance and differences in choice RTs across Goals conditions in Experiments 1-2 and 5-6. ]{\textbf{Correlations between differences in performance and differences in choice reaction times (RTs) across conditions of Goals blocks in Experiments 1-2 and 5-6.} Dots represent individual participants. Shading represents 95\% confidence intervals. Spearman's $rho$ correlation coefficients and p-values are shown on each respective plot.}
\label{fig:choice_rt_perf_corrs_by_group}
\end{figure}

\begin{figure}[t!]
\begin{center}
\includegraphics[width=0.6\textwidth]{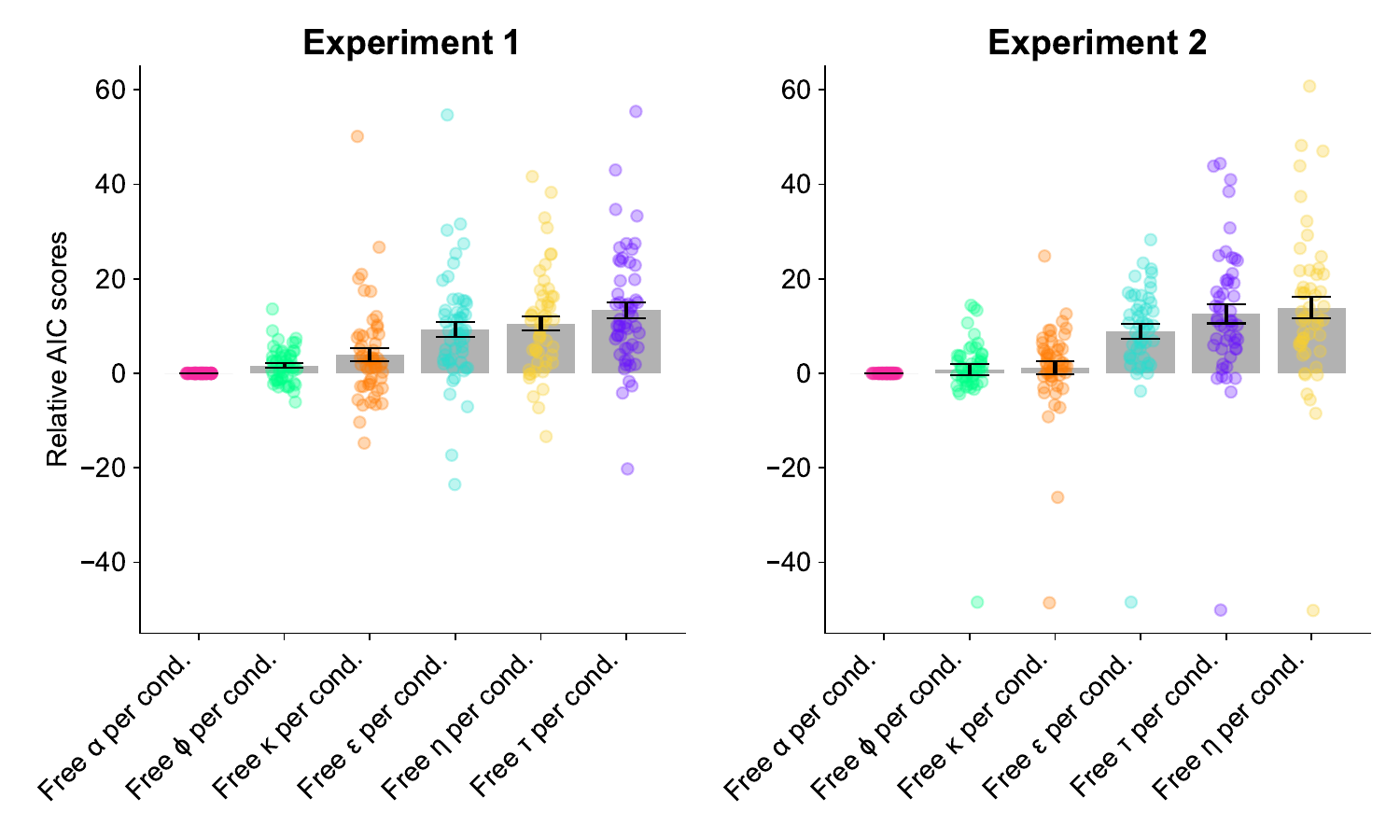}
\end{center}
\caption[Model comparison for Experiments 1 and 2 for models with free parameters for separate Goal sub-conditions.]{\textbf{Model comparison for Experiments 1 and 2 for models with free parameters for separate Goal sub-conditions.} Bars illustrate the average AIC score relative to the best model's across all participants. Dots represent relative AIC scores for individual participants. cond = condition.}
\label{fig:model_comparison_free_G_params}
\end{figure}

\clearpage
\begin{figure}[t!]
\begin{center}
\includegraphics[width=0.6\textwidth]{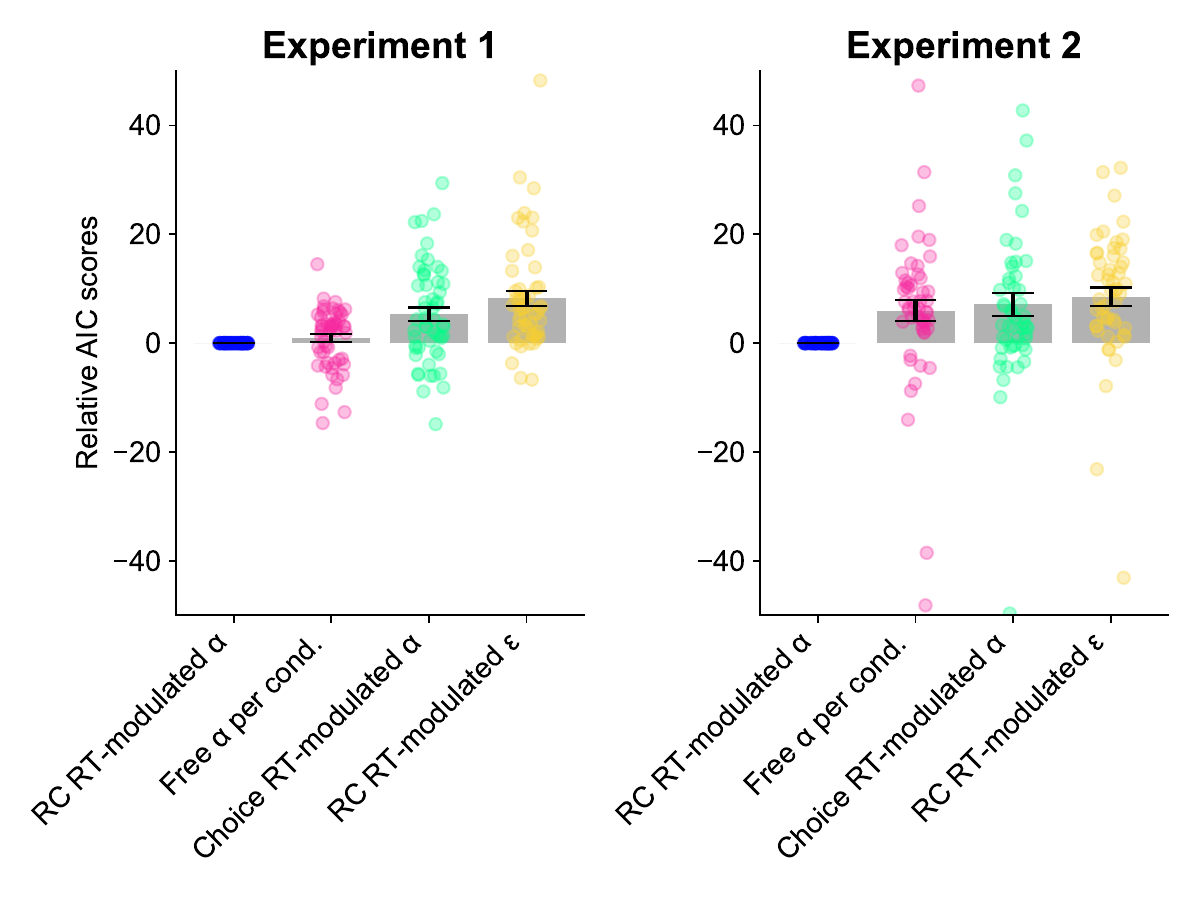}
\end{center}
\caption[Model comparison for the main models.]{\textbf{Model comparison for Experiments 1 and 2 for the main models discussed in the article.} Bars illustrate the average AIC score relative to the best model's across all participants. Dots represent relative AIC scores for individual participants. RC = reward collection. RT = reaction time.}
\label{fig:model_comparison}
\end{figure}

\clearpage
\begin{figure}[t!]
\begin{center}
\includegraphics[width=0.65\textwidth]{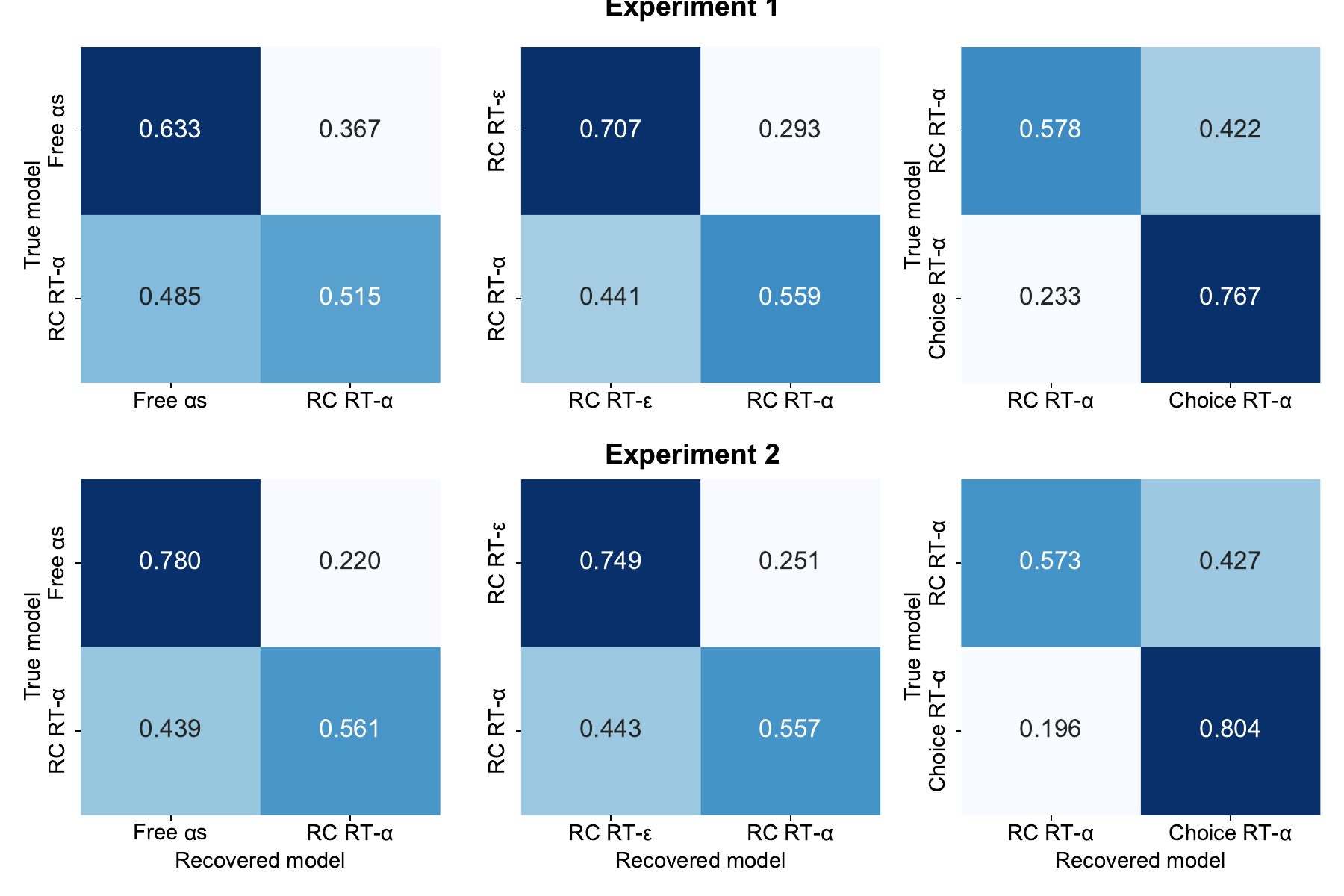}
\end{center}
\caption[Model recovery confusion matrix.]{\textbf{Model recovery confusion matrix.} We simulated behavior from each model 5 times per set of parameters and attempted to recover the best-fit model as measured by AIC. Each cell indicates the proportion of participants whose best-fit model corresponded to the true data-generation model. The top row illustrates results for Experiment 1, the bottom row for Experiment 2. The winning model, with learning rate $\alpha$ conditioned on reward collection reaction times (RC RTs; in ms) was compared to a model with free learning rates $\alpha$ (left plots), a model with noise parameter $\epsilon$ (right plots), and a model with $\alpha$ conditioned on choice RTs, separately. In all cases, recovery of our winning model RC RT-$\alpha$ was most likely when data was generated from that model than vs. the competing model.}
\label{fig:model_recovery}
\end{figure}

\clearpage
\begin{figure}[t!]
\begin{center}
\includegraphics[width=0.8\textwidth]{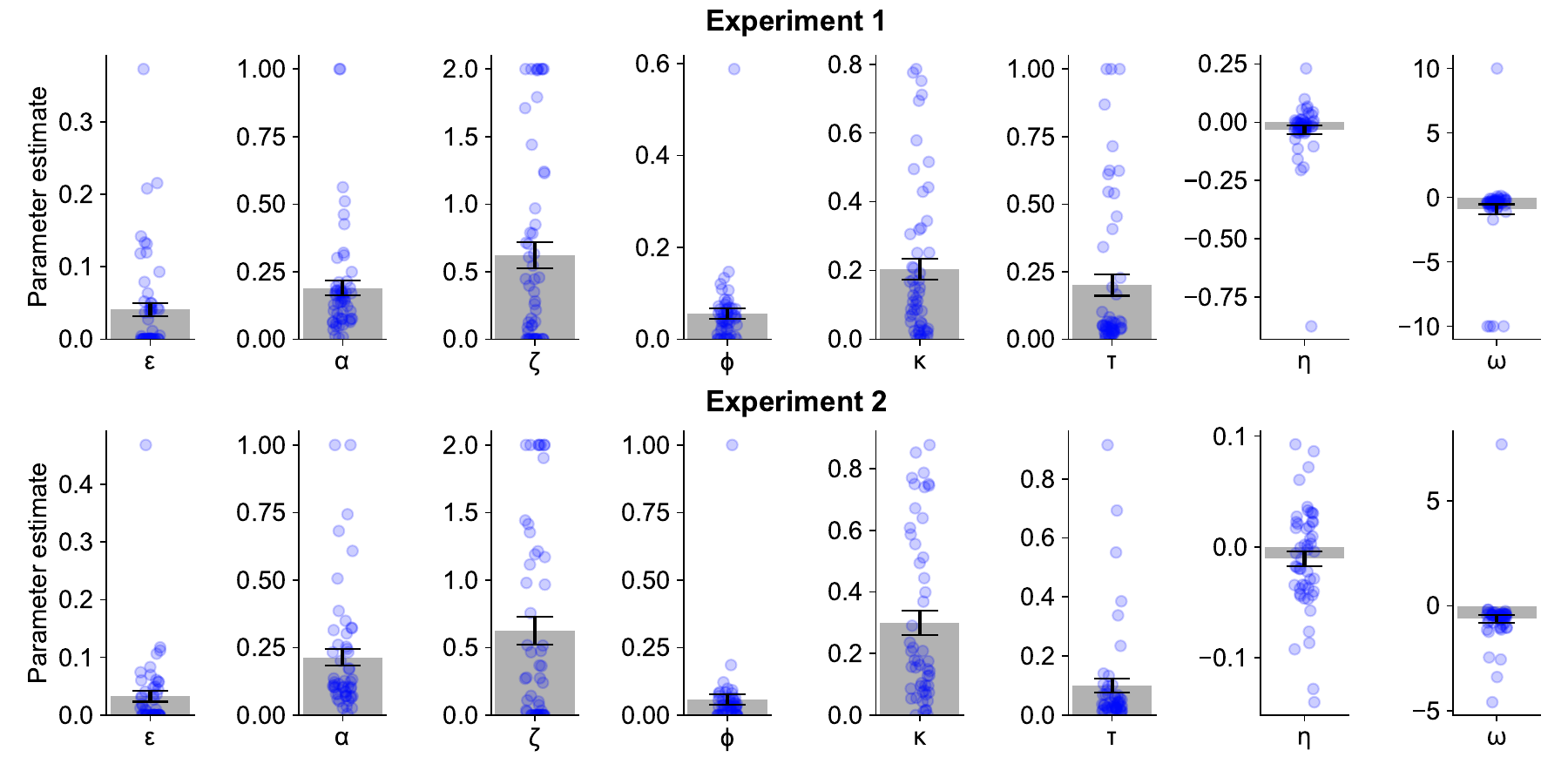}
\end{center}
\caption[Best-fit model parameters for the winning model.]{\textbf{Best-fit model parameters for the winning model, with learning rate $\alpha$ modulated by reward collection reaction times as determined by the weighting parameter $\omega$.} The average value of $\omega$ was significantly below zero, indicating that learning rates decreased with increased reward collection reaction times (Experiment 1: M =-0.94 $\pm$ 0.4, Z(53) = 60, p $<$ 0.001; Experiment 2: M = -0.62 $\pm$ 0.2, Z(50) = 51, p $<$ 0.001). Bars indicate parameter estimate averages. Error bars indicate the S.E.M. Dots represent individual participants' parameter estimates.}
\label{fig:best_fit_params}
\end{figure}

\clearpage
\begin{figure}[t!]
\begin{center}
\includegraphics[width=1\textwidth]{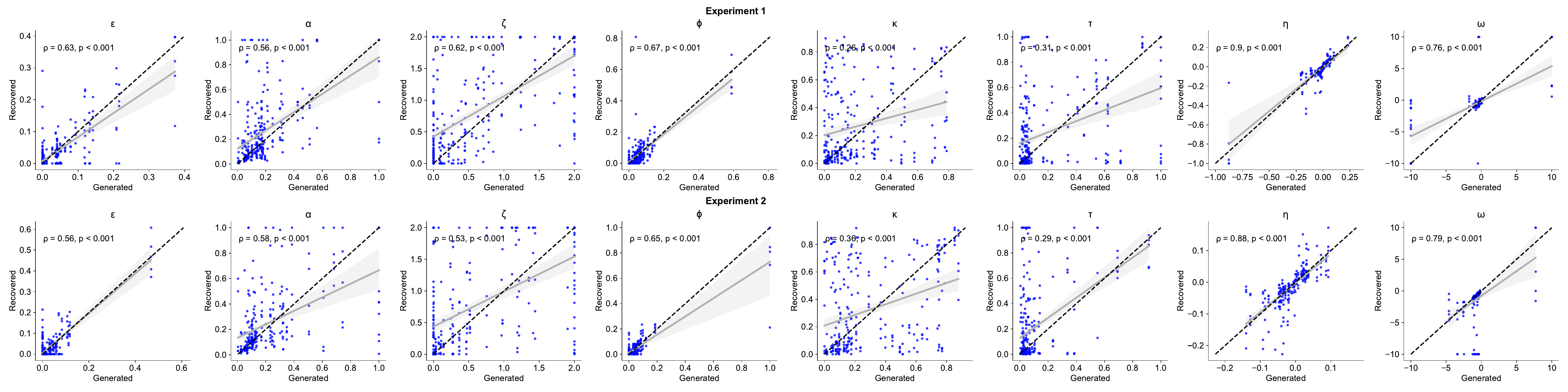}
\end{center}
\caption[Parameter recovery for the best model.]{\textbf{Parameter recovery for the best model.} Each plot shows the correlation between a seed parameter, i.e., the best-fit parameter for each participant -- used to simulate behavior -- and the parameter recovered through our fitting procedure. Each parameter was used for 5 separate iterations. Dots indicate individual parameters. Spearman's $\rho$ correlation coefficients and p-values are shown on each respective plot.}
\label{fig:param_recovery}
\end{figure}

\clearpage
\begin{figure}[t!]
\begin{center}
\includegraphics[width=1\textwidth]{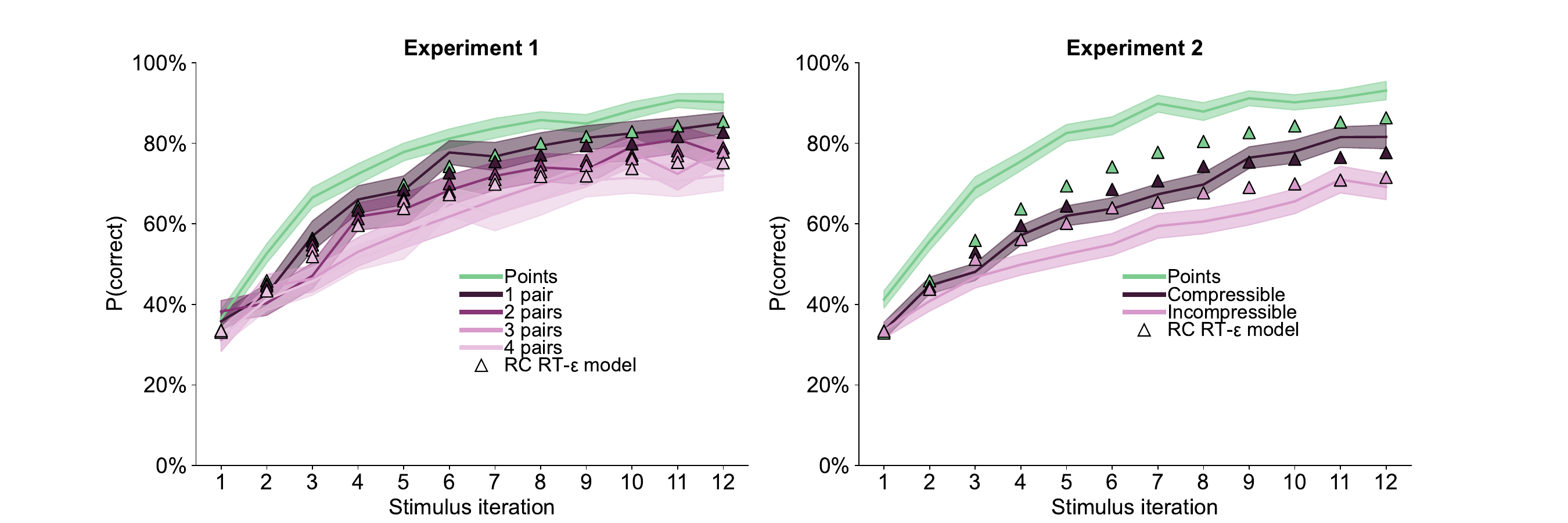}
\end{center}
\caption[Model validation with RT-modulated $\epsilon$.]{\textbf{Validation for a model with $\epsilon$ (choice noise) modulated by reward collection reaction times (RTs) for Experiments 1 and 2.} Solid lines and shading indicate participants' average data and the S.E.M., triangles show model estimates.}
\label{fig:alt_model_validation}
\end{figure}

\clearpage
\begin{figure}[t!]
\begin{center}
\includegraphics[width=1\textwidth]{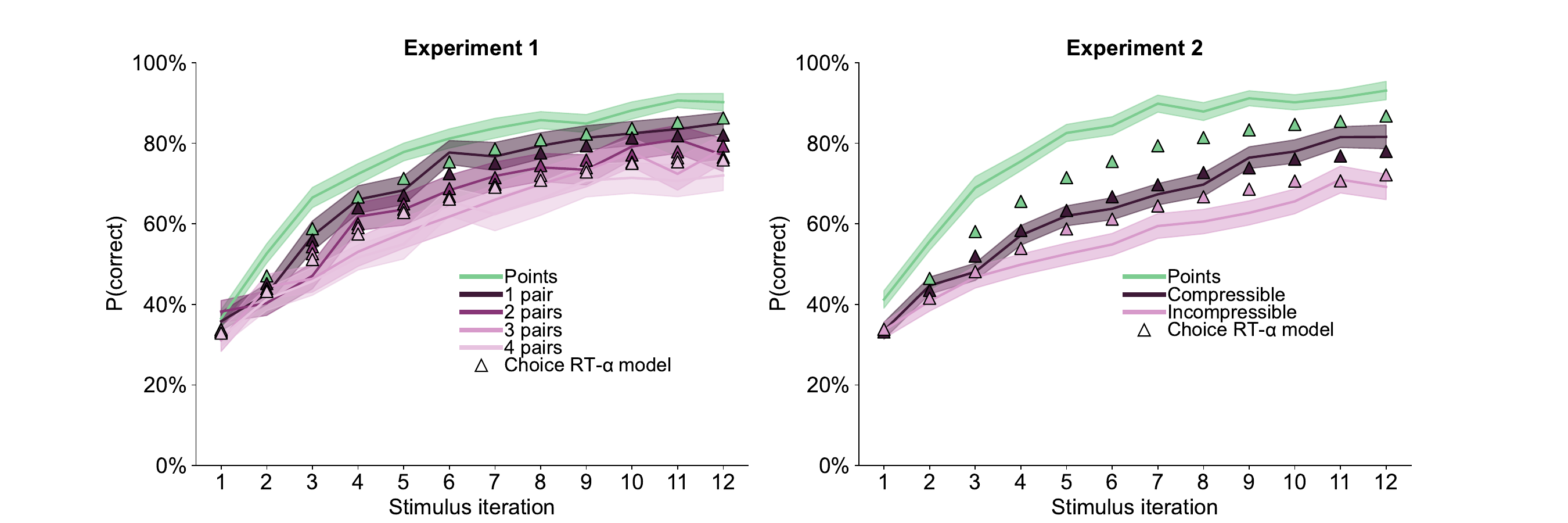}
\end{center}
\caption[Model validation with RT-modulated $\alpha$.]{\textbf{Validation for a model with $\alpha$ (learning rate) modulated by choice reaction times (RTs) for Experiments 1 (left) and 2 (right).} Solid lines and shading indicate participants' average data and the S.E.M., triangles show model estimates.}
\label{fig:choice_rt_model_validation}
\end{figure}

\stopcontents[mocked]
\stoplist[mockedfigures]{lof}

\end{document}